\documentclass[aps, pre, twocolumn, amsmath, superscriptaddress,showkeys,showpacs]{revtex4-1}
\usepackage{graphicx}
\usepackage{color, soul}
\usepackage{dcolumn,amsfonts}
\usepackage{bm}
\usepackage{epstopdf}
\usepackage{caption}
\usepackage{subcaption}
\usepackage{float}
\usepackage{ragged2e}
\usepackage{natbib}
\usepackage{hyperref}
\DeclareCaptionJustification{justified}{\justifying}
\begin{document}
\title{Role of assortativity in predicting burst synchronization using echo state network}
	
\author{Mousumi Roy}
\affiliation{Department of Applied Mathematics, University of Calcutta, 92, A.P.C. Road, Kolkata 700009, India}
\author{Abhishek Senapati}
\affiliation{Center for Advanced Systems Understanding (CASUS), 02826 G\"{o}rlitz, Germany}
\author{Swarup Poria}
\affiliation{Department of Applied Mathematics, University of Calcutta, 92, A.P.C. Road, Kolkata 700009, India}
\author{Arindam Mishra}
\email{arindammishra@gmail.com}
\affiliation{Division of Dynamics, Lodz University of Technology, Stefanowskiego 1/15, 90–924 Lodz, Poland}
\author{Chittaranjan Hens}
\affiliation{Physics and Applied Mathematics Unit, Indian Statistical Institute, Kolkata 700108, India}
\begin{abstract}
 In this study, we use a reservoir computing based echo state network (ESN) to predict the collective burst synchronization of neurons. Specifically, we investigate the ability of ESN in predicting the burst synchronization of an ensemble of Rulkov neurons placed on a scale-free network. We have shown that a limited number of nodal dynamics used as input in the machine can capture the real trend of burst synchronization in this network. Further, we investigate on the proper selection of nodal inputs of degree-degree (positive and negative) correlated networks.  We show that for a disassortative network, selection of different input nodes based on degree has no significant role in machine's prediction. However, in the case of assortative network, training the machine with the information (i.e time series) of low-degree nodes gives better results in predicting the burst synchronization. Finally, we explain the underlying mechanism responsible for observing this differences in prediction in a degree correlated network.
 
\end{abstract}
\maketitle

\section{Introduction}

Reservoir computing (RC) is a simple yet extremely efficient machine learning methodology for predicting temporal data. There are several techniques, e.g., Echo state network (ESN), liquid state machine (LSM), those have been developed depending on the mechanism of reservoir computing. Since the seminal work by Jaeger \textit{et al.}~\cite{jaeger2001echo,jaeger2004harnessing}, it has opened up a new direction in the field of ESN based machine learning approach~\cite{lukovsevivcius2009reservoir,rodan2010minimum,farkavs2016computational,koryakin2012balanced,tanaka2019recent,lukovsevivcius2012practical}. Due to its simplicity, ESN is appeared as a considerable tool in different areas ranging from neuroscience~\cite{kim2019decoding,hramov2021physical}, speech recognition~\cite{skowronski2007automatic}, language processing~\cite{hinaut2013real}, robotics~\cite{chessa2014robot} to stock market prediction~\cite{lin2009short}, inference of connectivity~\cite{banerjee2019using,banerjee2021machine}, network classification~\cite{panday2021machine} and even in predicting recent COVID-19 epidemic trends~\cite{ghosh2021reservoir}. There are also some studies which have focused {on the suitable choices of   reservoir
weights}~\cite{rodan2010minimum,qiao2016growing,cui2014effect} and on the optimal range of hyper parameters~\cite{venayagamoorthy2009effects,buhlmann2003boosting,platt2021forecasting,griffith2019forecasting,lu2018attractor,kawai2019small,Alexander2019,lymburn2019consistency,carroll2019network,shirin2019stability}, which are crucial for a good prediction. {On the other hand, despite the improvement of diverse nonlinear and statistical tools \cite{kantz2004nonlinear,parlitz2000prediction}, the understanding of ergodicity  as well as analyzing the chaotic feature pose challenges for the researchers in the field of nonlinear dynamics.

\par 
 In this backdrop, it has been shown that ESN based machine learning approach can be highly useful for the efficient prediction of nonlinear data due to its simple architecture and faster computation.    For instance,} Pathak \textit{et al.}~\cite{pathak2018model} showed that a reservoir computation based ESN approach can indeed predict a large spatiotemporal chaotic data and the prediction efficiency is very good up to few multiples of Lyapunov time  scale.
Since then, ESN based prediction has received  growing attention from the researchers and has been used in several aspects of nonlinear dynamics, for example, prediction of a chaotic time series data~\cite{weng2019synchronization,borra2020effective,pathak2018model,maslennikov2019collective,zhang2020predicting}, spatiotemporal dynamics~\cite{zimmermann2018observing}, determining Lyapunov exponents~\cite{pathak2017using,carroll2018using}, dynamics of multiscale systems~\cite{borra2020effective}, predicting critical transition~\cite{kong2021machine} and critical range for the efficiency of the reservoir~\cite{mandal2021achieving}, to name a few. Another interesting aspect namely the collective or macroscopic behavior of ensemble of interacting oscillators, such as, synchronization, quenching of oscillations, chimera states etc. \cite{pikovsky2003synchronization,arenas2008synchronization,koseska2013oscillation,saxena2012amplitude,parastesh2020chimeras} {, those can be efficiently identified and extracted with the machine learning tools. The critical parameter for the amplitude death ~\cite{xiao2021predicting}, and the  onset of generalized synchronization can  easily be captured using ESN~\cite{lymburn2019reservoir,ibanez2018detection} based approach.} First-order and second-order phase transition of a system of non-identical oscillators can be predicted through ESN as well~\cite{fan2021anticipating}. There are also other approaches of machine learning for identification of chimera and solitary states~\cite{ganaie2020identification,kushwaha2021machine}, generalized synchronization~\cite{frolov2019feed}, predicting extreme events~\cite{chowdhury2021extreme,ray2021optimized} and forecasting COVID-19 spread~\cite{chakraborty2020real}, but, we consider ESN for our study for its less computational cost.  
\par 
 Motivated by  such diverse applications of ESN in nonlinear systems, we target here to predict the collective spiking and bursting dynamics in a network of non-identical neurons. Particularly,  we develop a  modified architecture of ESN to capture the onset as well as the strength of  burst synchronization. It is to be noted that spiking is a repeated firing state of neurons, whereas bursting is a bunch of spikes  \cite{izhikevich2000neural,hens2015bursting,mishra2021neuron,roy2021assortativity,rulkov2002modeling,ghosh2020emergence}. Each  burst is followed by a  quiescence state before the next burst occurs and it can be periodic or chaotic. In a recent study~\cite{saha2020predicting}, it has been revealed that for a mixed population of coupled oscillators, ESN needs at least one time series data from each group to get a satisfactory spiking and bursting prediction as well as predicting the onset of generalized synchronization. However, this prediction of bursting dynamics by ESN {was limited to a complete (uncorrelated) graph and the original as well as machine generated profile of the bursts were periodic. Also, a slight increase of the coupling makes the system into two tight cluster states and machine  predicts the entire time signals  more accurately. However, broad spectrum of randomness in the system parameters (desynchronized chaotic neurons) will not generate cluster synchronization immediately after the coupling is active, rather the  phases of neurons  may start to synchronize earlier \cite{roy2021assortativity}. To identify such phase synchronization (particularly starting and ending of the bursts of each oscillator at the same time) motivates us to use the machine learning(ML): whether ESN can predict the burst synchronization for non-identical neurons. On the other hand, the network architecture encoded in neuronal connection is not regular rather follows a complex structure \cite{eguiluz2005scale,zhou2006hierarchical,bassett2006small,bassett2006adaptive}. Thus we target here to predict the burst synchronization of neurons connected in complex network through the modified approach of ESN. To delve deeper, we have investigated the emergence of bursts in an assortative as well as disassortative networks~\cite{newman2002assortative,newman2003mixing}. Assortative mixing is associated with several collective events, such as, functional network in human cortex~\cite{eguiluz2005scale}, spreading phenomena through complex network~\cite{yang2015traffic,schlapfer2012decelerated}, information networks~\cite{pastor2001dynamical}, etc. Interaction among similar individuals are subjected to network resilience, makes a system more stable under external perturbations.
In this regard, the key questions we have raised here what would be the minimum number of input neurons required for an efficient prediction and depending on the level of assortaivity of the original system which nodes should be used as the inputs to the machine? 
%
%

\par
As an attempt to address theses questions, we concentrate on making an ESN based prediction of the degree of burst synchronization of a neuronal system on a complex network with positive and negative degree-degree correlation. We consider an ensemble of Rulkov neurons, diffusively coupled though a heterogeneous scale-free network. It is well known that the system attains a phase synchronized state through a continuous phase transition process~\cite{batista2007chaotic}. At this stage, in an ESN environment, we train the machine using time series data of a few nodes to acquire the predicted time series of remaining nodes. Here, in stead of whole time series, we focus on the onset of bursts and try to predict the route of the synchronization transition process. Furthermore, we explore whether any particular choice of input nodes results a better prediction. In this context, since assortative mixing  is a local network characteristic having a significant impact in collective dynamics of the system, we try to use this characteristic as a base of selecting input nodes to train the machine. Surprisingly, this aspect of finding an optimal choice of input nodes is still not well explored. Therefore, in this study, we try to elucidate how degree dependent assortativity of the original system makes a discrepancy in the predicted results. Moreover, we also examine how we can improve the results using different group of training nodes. Finally, we explain these findings based on the dynamics of original system. \par
Here, we attach a brief outline of this study. In Sec.~\ref{sec:esn}, a detailed description of the reservoir system is presented. We briefly describe the Rulkov map based model and structure of the underlying complex network in Sec.~\ref{sec:rulkovmodel}. Sec.~\ref{sec:quantification} is devoted to define some quantities like assortativity, synchronization order parameter, root mean square error, etc. that are used throughout the study in quantifying the efficiency of the prediction. Next we organize all the results in Sec.~\ref{sec:results}, and end up with a conclusion in Sec.~\ref{sec:conclusion}.

\section{ESN architecture}
\label{sec:esn}
In this study, we consider a conventional ESN, a discrete time leaky {\it{tanh}} system updates its internal dynamics using the formula:
\begin{equation}
v(t+ 1)=(1-\alpha)v(t)+\alpha \times \tanh\begin{pmatrix}W_{\rm res}v(t)+W_{\rm in}\begin{bmatrix}1 \\u(t)\end{bmatrix}\end{pmatrix}
\label{reservoir_dynamics}
\end{equation}
where, $W_{\rm res}(\in\mathbb{R}^{N_{\rm res}\times N_{\rm res}})$ is the reservoir matrix. It is chosen as a sparse matrix that determines the connectivity weights into the reservoir. It takes the non zero elements randomly chosen from $[-1,1]$. We fix the spectral radius of $W_{\rm res}$ at $\rho=0.95$ and rescale the matrix accordingly (we consider the spectral radius as less than unity for preventing the reservoir to settle down on multiple fixed points, periodic or even chaotic attractor modes, thus ensuring the echo state property). $v (\in\mathbb{R}^{N_{\rm res}})$ represents the dynamics of internal reservoir nodes. \par
{Here, we use a system of Rulkov neurons to investigate the efficiency of the ESN generated prediction process. A detailed description of the model is given in Sec.~\ref{sec:rulkovmodel}. From this system, to train ESN we use the time series of $K$ chosen nodes for first $t_{train}$ time points (after ignoring a considerable transient time), which is the term $u(t)$ in Eq.~\ref{reservoir_dynamics}, defined as,}
 $$u(t)=\begin{bmatrix}u_1(t)\\u_2(t)\\.\\.\\.\\u_K(t)\end{bmatrix}.$$ 
$W_{\rm in}(\in\mathbb{R}^{N_{\rm res}\times (K+1)})$ is the corresponding weights to connect the inputs with reservoir. We randomly take the elements of $W_{in}$ from the uniformly distributed interval [-0.5,0.5]. $\alpha$ is the leaking parameter can take values from $0$ to $1$. Particularly in this study, we consider $\alpha=0.09$ and $N_{\rm res}=1000$.  \par
Once the training is done, our target is to predict the dynamics of remaining $N-K$ nodes from the machine generated time series for next $t_{\rm test}$ time points. During the training period, at each time instant the readout weights are collected into a matrix $X$, where $X (\in \mathbb{R}^{(N_{\rm res}+K+1)\times t_{\rm train}})$ is defined as,\\
\begin{equation*}
X=\begin{bmatrix}1 & 1 &...&1\\
u_1(1)&u_1(2)&...&u_1(t_{\rm train})\\
u_2(1)&u_2(2)&...&u_2(t_{\rm train})\\
. & .&.&.\\
. & .&.&.\\
. & .&.&.\\
u_K(1)&u_K(2)&...&u_K(t_{\rm train})\\
v_1(1)&v_1(2)&...&v_1(t_{\rm train})\\
v_2(1)&v_2(2)&...&v_2(t_{\rm train})\\
. & .&.&.\\
. & .&.&.\\
. & .&.&.\\
v_{N_{\rm res}}(1)&v_{N_{\rm res}}(2)&...&v_{N_{\rm res}}(t_{\rm train})\end{bmatrix}.
\end{equation*}
This gives us the system of linear equations of readout weights,
\begin{equation}
Y=W_{\rm out}X,
\label{equation2}
\end{equation}
where, $Y(\in \mathbb{R}^{(N-K)\times t_{\rm train}})$ is the time series of the targeted $N-K$ nodes for the training period$(t=1,2,3,...,t_{\rm train})$,
Therefore, we can find $W_{\rm out}(\in \mathbb{R}^{(N-K)\times (N_{\rm res}+K+1)})$ as 
\begin{equation}
W_{\rm out}=YX^t(XX^t+\lambda I)^{-1}, 
\end{equation}
$\lambda$ is the ridge regression parameter to avoid over fitting during the training process.\\
Now, for testing we obtain $Y_{\rm target}$ by employing $W_{\rm out}$ on $X$ at next $t_{\rm test}$ time points. First we use $u(t)$, time series of the initially chosen $K$ input nodes for $t=t_{\rm train+1},t_{\rm train+2},...,t_{\rm test}$ into Eq.~\ref{reservoir_dynamics}. At every time instant we get $v(t)$ as well as $X(t)=[1;u(t);v(t)]$, and  using Eq.~\ref{equation2}, we finally obtain the targeted result
$$Y_{\rm target}=W_{\rm out}X, \hspace{0.3cm} \text{for} \hspace{0.3cm} t=t_{\rm train+1},t_{\rm train+2},...,t_{\rm test}.$$

\section{Dynamics of coupled Neuron Model}

\label{sec:rulkovmodel}
We consider an ensemble of diffusively coupled $N$ chaotic Rulkov neurons~\cite{rulkov2002modeling}. At every time instant the system follows the equations :
\begin{equation}
x_i(t+1)=\frac{\alpha_{i}}{1+x_i^2(t)}+y_i(t)+\epsilon\sum_{j=1}^N C_{ij} (x_j(t)-x_i(t)) ,
\label{rulkov1}
\end{equation}
\begin{equation}
y_i(t+1)=y_i(t)-\sigma(x_i(t)-\beta),         \hspace{0.8cm} i=1,2,3,...,N.
\label{rulkov2}
\end{equation}
 where, $\alpha_{i}$ is randomly chosen from the uniform distribution $[4.1,4.5]$. We take $\beta=-1.5$ and $\sigma=0.003$. For this parameter set, isolated node follows a chaotic bursting dynamics (as shown in Fig.~\ref{Fig:phase}). Here $\epsilon$ is the coupling strength and $C$ is the adjacency matrix of a Barab\'{a}si–Albert(BA) scale free network~\cite{barabasi2009scale}. For this study, we consider $N=500$ and average degree $6$.\par

 \begin{figure}
	\begin{center}
		{\includegraphics[width= 0.5
			\textwidth]{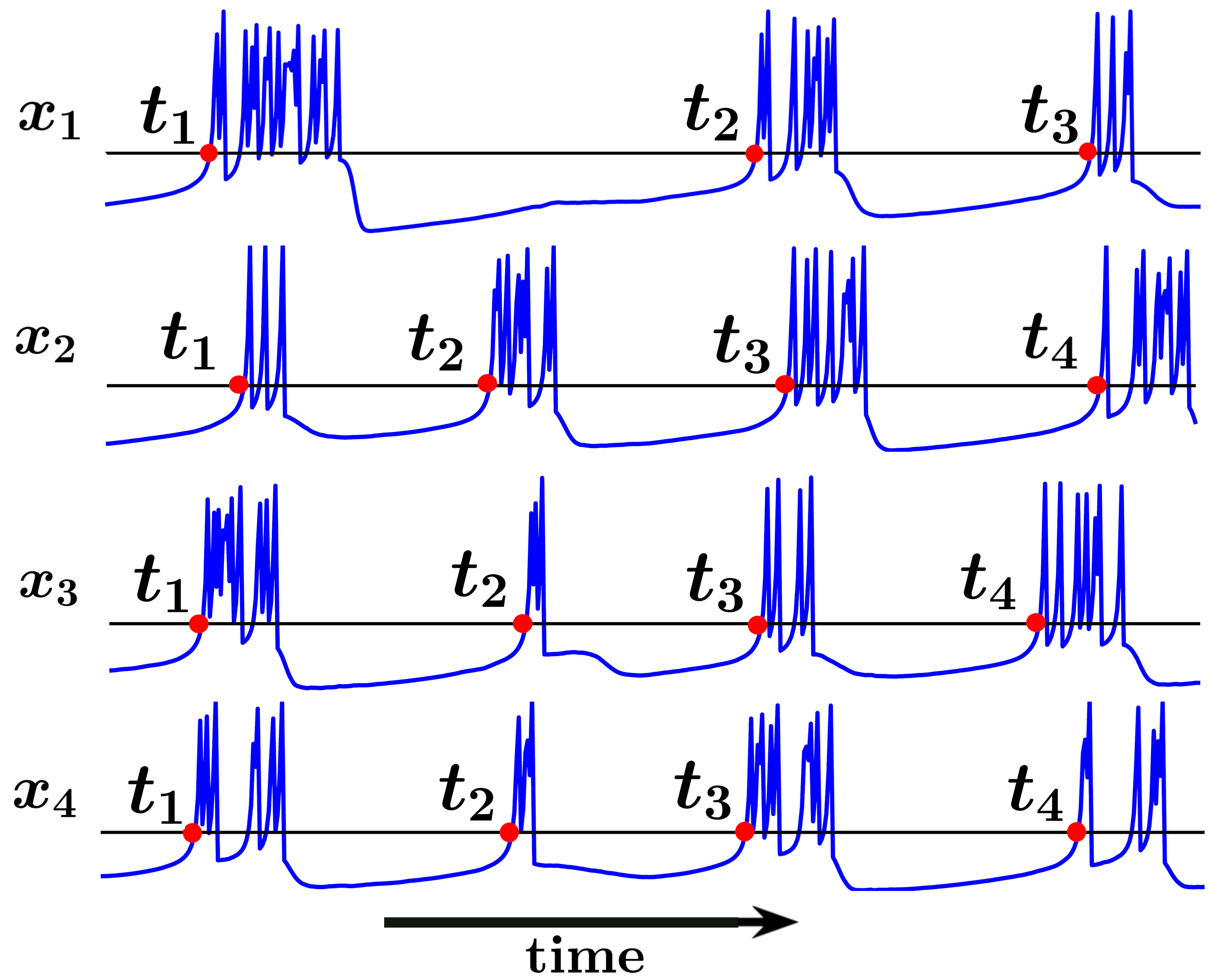}} 
	\end{center}
	\caption{\textbf{Time series of four randomly chosen nodes of the uncoupled system.} Individual node follows the chaotic bursting dynamics and every horizontal black lines are the Poincar\'{e} section $x_0=-1$. Red dots are the starting points of every burst those are used for calculating phases.}
	\label{Fig:phase}
\end{figure}

\section{Uncorrelated network: Results and quantification}
\label{sec:quantification}
Initially, we consider the original system~\ref{rulkov1}-\ref{rulkov2} connected through an uncorrelated network with assortativity $A=0$. Assortativity $A$ is defined as the degree-degree Pearson correlation as follows:
 \begin{equation}
 A=\frac{M^{-1}\sum_ij_ih_i-[M^{-1}\sum_{i}\frac{1}{2}(j_i+h_i)]^2}{M^{-1}\sum_i\frac{1}{2}(j_i^2+h_i^2)-[M^{-1}\sum_i\frac{1}{2}(j_i+h_i)]^2},
 \label{assortativity}
 \end{equation}
 where, $j_i$ and $h_i$ are degrees of the nodes connected through $i^{th}$ link, and $M$ is the total number of links associated with the network.
 Next, we train the machine using the time series of $t_{\rm train}=25000$ data points (avoiding first $20000$ transient time points) of  $K$ chosen nodes. Our target is to predict the time series of remaining $N-K$ nodes for next $t_{\rm test}=10000$ time instants. In Fig.~\ref{Fig:mean_field}, we plot the time series of a randomly chosen nodes (see Figs.~\ref{Fig:mean_field}(a)-(d)) and the corresponding mean fields (see Figs.~\ref{Fig:mean_field}(e)-(h)) of whole system from both the original and machine predicted data at several coupling strengths $(\epsilon= 0.02, 0.04, 0.06, 0.12)$ where mean field is defined as,
$$X(t)=\frac{1}{N}\sum_{j=1}^{N}x_j(t).$$
Here, we randomly choose $K=15$ input nodes to train the machine. We use the colors gray and red to present the original and machine predicted data respectively. Interestingly, we observe that machine is not able to predict individual time series with a perfect amplitude matching (see Figs.~\ref{Fig:mean_field}(a)-(d)) but the mean field (see Figs.~\ref{Fig:mean_field}(e)-(h)) and the onsets of bursts are considerably well captured. At considerably low coupling strength, $(\epsilon=0.02)$, machine's prediction is very poor but with the increment of coupling strength an improvement in the predicted result is noticeable. For comparatively higher coupling strength ($\epsilon=0.06,0.12)$, even though exact amplitude prediction has not been achieved, machine can accurately produce mean field of the whole system (see Figs.~\ref{Fig:mean_field}(g),(h)). We see that instead of individual time series, machine can capture the onset of bursts and as well as overall mean field dynamics quite well at various coupling strengths, that motivates us to focus on predicting the degree of burst synchronization of the whole system. \par

  \begin{figure*}
	\begin{center}
		{\includegraphics[width= 1
			\textwidth]{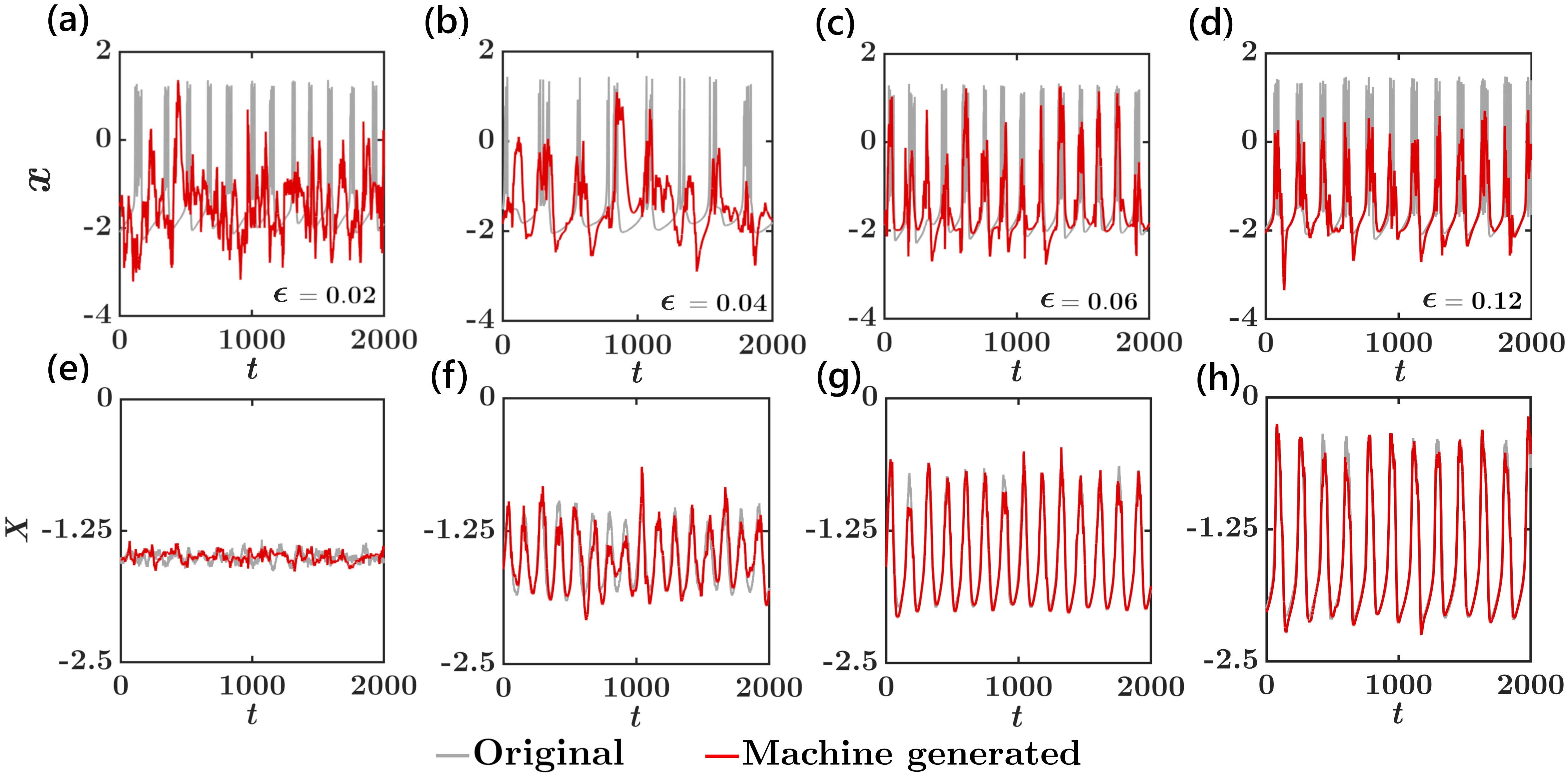}}
	\end{center}
	\caption{\textbf{((a)-(d)) The original and machine generated time series of a randomly chosen node at $\epsilon=0.02,0.04, 0.06, 0.12$ and the corresponding time series of mean field are represented in the lower row ((e)-(h)) respectively.} ESN is unable to predict the individual time series but it can capture the overall dynamics of the mean field quite efficiently for coupling strength $\epsilon \geq 0.04$. Here, we consider a scale-free network with total number of nodes $N=500$ and average degree $6$. }
	\label{Fig:mean_field}
\end{figure*}


 As we can see that machine can not capture the individual amplitudes of the oscillators, we have to rely on phases in order to predict burst synchronization. We measure the phase $\theta$ from the time series of every neuron by introducing a Poincar\'{e} section at $x_0=-1$. Every time the time series of $x$ crosses the line $x_0=-1$ in upward direction we take those points as the starting points of new bursts (see Fig.~\ref{Fig:phase}). We also put a restriction that two simultaneous bursts should apart by at least 60 time points.Then, mathematically, phase of a chaotic oscillator $\theta_j(t)$ is defined as~\cite{pikovsky2003synchronization},
 \begin{equation*}
 	\theta_j(t)=2\pi n+2\pi\frac{t-t_{n,j}}{t_{n+1,j}-t_{n,j}} , \hspace{1cm} t_{n,j}\leq t< t_{n+1,j},
 \end{equation*}
 where, $t_{n,j}$ is the starting time point of $n^{th}$ burst of $j^{th}$ neuron. in Fig.~\ref{Fig:phase}, we present the time series of four randomly chosen nodes from the uncoupled system, and those red dots find the starting point of a new burst. We take the data after avoiding a considerable transient time points.  \par
Now, with the phases of the oscillators, we  can quantify the degree of burst synchronization using the absolute value of Kuramoto's order parameter~\cite{kuramoto1991collective} as,
\begin{eqnarray}
r(t)=\left|\frac{1}{N}\sum_{j=1}^{N}e^{i\theta_j(t)} \right|,
\label{kuramoto}
\end{eqnarray}
where, $\theta_j(t)$ is the phase of $j^{th}$ neuron at time $t$. For this calculation, we separately use the original and machine generated time series for the $t_{\rm test}$ time points. Averaging \eqref{kuramoto} over a large time interval, we get the final averaged out order parameter
\begin{eqnarray}
 R =\frac{1}{t_f-t_i}\sum_{t=t_i}^{t_f}r(t),
 \label{eq:kuramoto order parameter}
\end{eqnarray}
where, $t_f$ is the final time point, and $t_i$ is the first time point after ignoring the transient period. This global synchronization order parameter $R\rightarrow1$ signifies a perfectly phase synchronized state and in contrast, $R\rightarrow0$ for an entirely non-phase synchronized state. \par

In order to quantify the order of a good prediction, we consider the values of root mean square error ($RMSE$) between the original and predicted time series of $r(t)$. $RMSE$ is calculated using the formula : 
\begin{equation}
RMSE=\sqrt{\frac{1}{t_f-t_i}\sum_{t=t_i}^{t_f}(r_{\rm original}(t)-r_{\rm machine}(t))^2}. 
\end{equation}\par
In Fig. ~\ref{Fig:A0_prediction}, we take the same uncorrelated network as in Fig.\ref{Fig:mean_field}, and plot the values of $R$ with respect to $\epsilon$ for different number of input nodes. The original (black circle) and machine predicted (green square) order parameters are represented for $K=5,15$ and $25$ (see Figs. \ref{Fig:A0_prediction}(a)-(c)). We randomly choose the input nodes and observe that with increasing values of $K$, the prediction improves gradually. In Fig.~\ref{Fig:A0_prediction}(d), corresponding RMSE justifies this results shown in Fig.~\ref{Fig:A0_prediction}(a)-(c). RMSE gradually decreases  with increasing values of $K$. Therefore, \textit{for an uncorrelated network, increasing inputs help the machine to generate an improved prediction. }  \par
But what will be the scenario if the network has a degree-degree correlation between the nodes and for this is there any preference of choosing specific input nodes? We address these questions in the next sections.

  \begin{figure*}
	\begin{center}
		{\includegraphics[width= 1
			\textwidth]{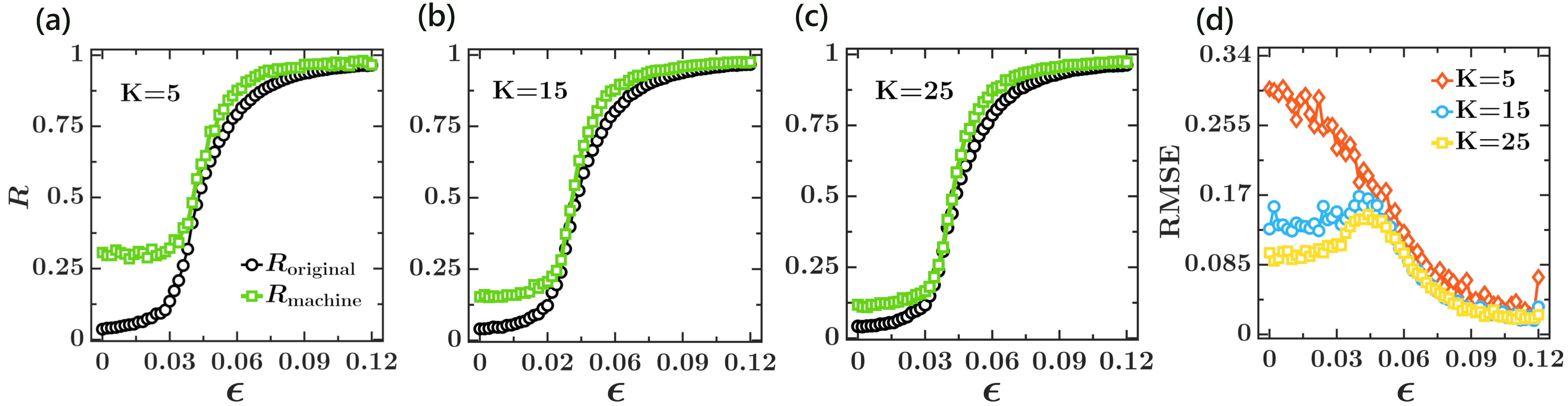}}
	\end{center}
	\caption{\textbf{Prediction of synchronization level through ESN.} ((a)-(c)) Global synchronization order parameter $R$ is represented with respect to $\epsilon$ from the original (black circle) and machine generated (green square) data for different values of $K$. Here, we take an uncorrelated network. Input nodes are randomly chosen from the whole system. (d) RMSE is depicted with respect to $\epsilon$. At low coupling strength, increasing inputs make a better prediction. }
	\label{Fig:A0_prediction}
\end{figure*}
\section{Correlated networks: Effect of positive and negative correlations}
\label{sec:results}
In this section, we address the questions on whether the types of nodes we select to train the machine have any effective influence in the predicted results and whether it has a connection with the assortativity of the original network. Assortativity ($A$) is one of the important characteristics of a scale-free network that quantifies the tendency of connecting similar degree nodes. We make a network assortative by removing links between high and low degree nodes and reconnect two similar degree nodes. On the other hand, by reversing this process in opposite sense, we can make a network disassortative. We follow the algorithm, proposed in~\cite{xulvi2004reshuffling} to get a network having  desired level of degree assortativity. It is measured by Pearson's correlation coefficient as described in equation~\ref{assortativity}. At different level of assortativity, local network topology reconstructed without affecting the degree distribution of the whole network. But this local rearrangement of links have an effective impact in global scenario. This leads us to investigate the effect of assortativity index over the input node selection process during the training of ESN.
Therefore, we consider networks with different level of assortativity and use different group of nodes for training. We consider the nodes with degree less than 6 as the low degree nodes, with degree greater than 15 as the high degree nodes and the intermediate nodes are having degree between 6 to 15.   \par
At a global synchronized state, considering large values of $K$ (number of input nodes) is absolutely unnecessary. Feeding the data of only one node is enough to predict the phases of the remaining nodes (since, they are all synchronized). Therefore, we concentrate on the desynchronized or partially synchronized state of the system i.e for the low coupling strength $(<\epsilon\sim0.06)$.
\subsection{Effect of disassortativity}
\begin{figure*}
	\begin{center}
		{\includegraphics[width= 1
			\textwidth]{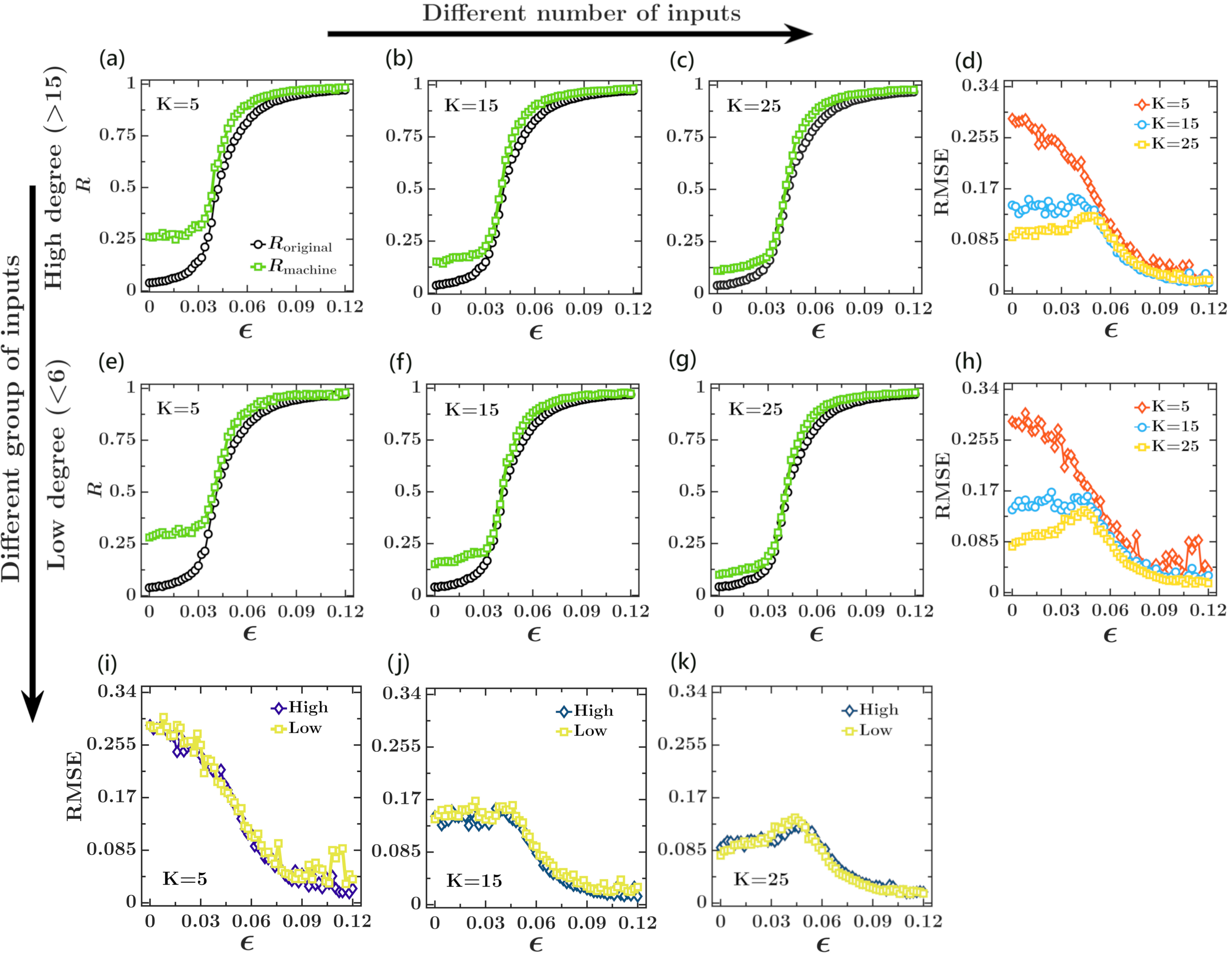}}
	\end{center}
	\caption{\textbf{A disassortative \boldmath$(A=-0.3)$ network is considered. The global synchronization order parameter $R$ from the original (black) and machine predicted (green) data is depicted with respect to $\epsilon$.} We choose the inputs from high ((a)-(c)) and low ((e)-(g)) degree groups. In the last column, we present RMSE with the variation of $\epsilon$ for different number of (d) high  and (h) low degree inputs and with the increment in the number of inputs we obtain better prediction. (i)-(k) RMSE for different group of input nodes. Selection of lower and higher degree nodes during the training do not show any distinct effect on the efficiency of prediction process.}
	\label{Fig:assort-0.3}
\end{figure*}
We start by considering a disassortative network (negatively correlated) with $A=-0.3$. In Fig.~\ref{Fig:assort-0.3}, we plot the values of $R$ from both original and machine generated time series data with respect to $\epsilon$ for different number of inputs. Here, we make a preference in choosing these training nodes depending on their degree, like high (degree$>15$) and low (degree$<6$) degree. In the upper panel (see Fig.~\ref{Fig:assort-0.3}(a)-(d)), the time series of high degree nodes are used to train the machine and, we gradually increase the number of input nodes $(K)$ from 5 to 25 (see Fig.~\ref{Fig:assort-0.3}(a)-(c)). We use black and green colors for the original and machine predicted data. It is observed that the increment in the number of inputs helps ESN to capture the synchronization level of the whole system. Fig.~\ref{Fig:assort-0.3}(d) justifies this observation as RMSE decreases with the increasing values of $K$. Here, red, blue and yellow colors are used for $K=5, 15$ and $25$. Increasing inputs favors in decreasing RMSE gradually at the desynchronized or partially phase synchronization state. In the lower panel (see Figs.~\ref{Fig:assort-0.3}(e)-(h)), the inputs are chosen from the low degree nodes, and we find similar results as the high degree inputs. \par
In the last row, we compare RMSE for different group of inputs at some fixed value of $K$. Blue and yellow colors are used for high and low degree groups, and we observe that RMSE takes almost similar values (see Figs.~\ref{Fig:assort-0.3}(i)-(k)) for these two different group of input nodes. This leads us to the fact that, {\it{for a disassortative network, depending on degree, preference in selecting different group of training nodes has no significant impact on the machine predicted results. Only increasing number of inputs helps ESN to make a better prediction as before for an uncorrelated network}}.     
\subsection{Effect of assortativity}
\begin{figure*}
	\begin{center}
		\includegraphics[width= 1\textwidth]{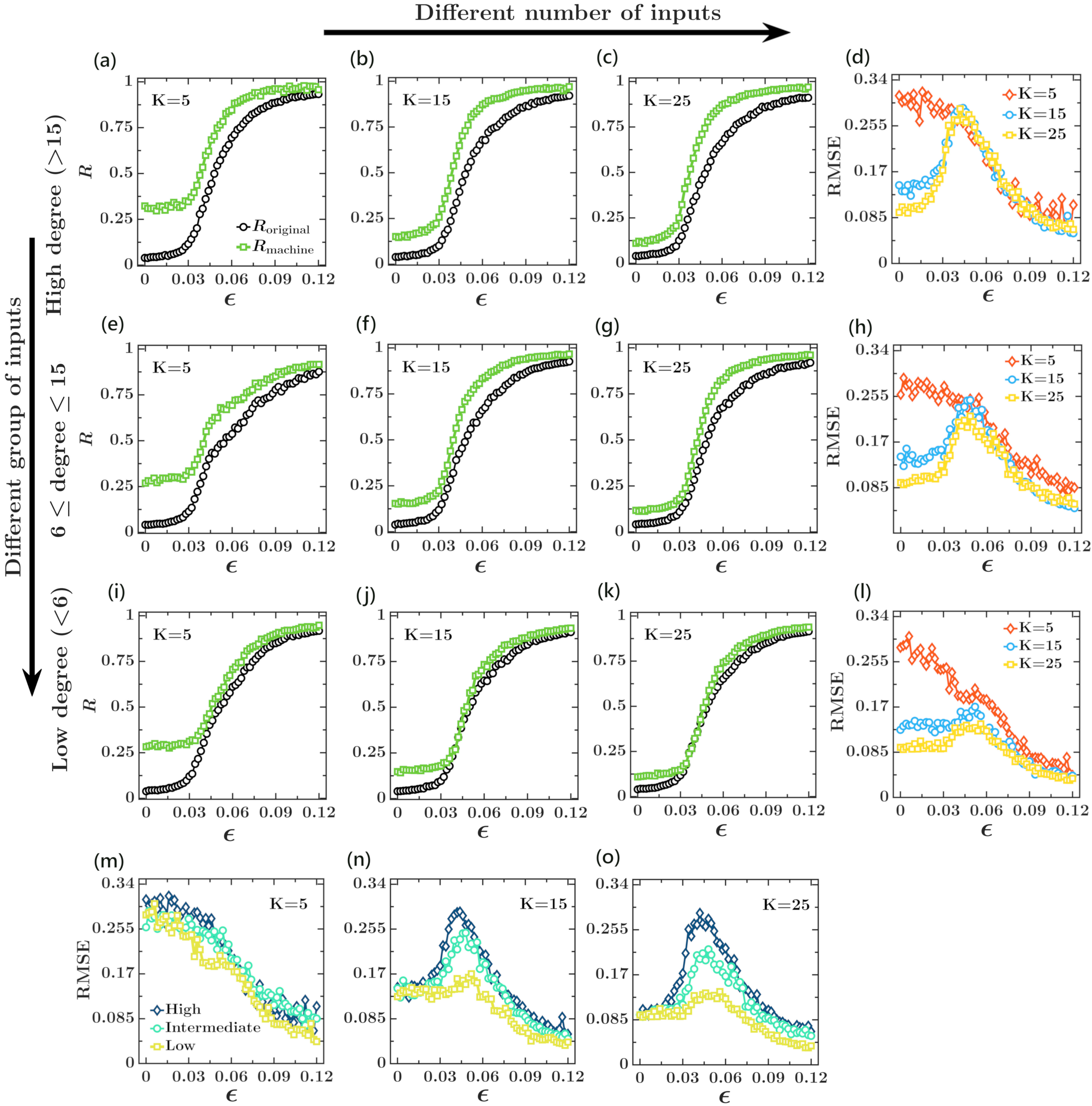}
	\end{center}
	\caption{\textbf{An assortative \boldmath$(A=0.3)$ network is considered. In the first, second, third row, we plot the values of $R$ from original (black) and machine generated (green) data with respect to $\epsilon$ for three different group of inputs.} High (a)-(c), intermediate (e)-(g) and low degree (i)-(k) nodes are used as inputs. RMSE for different number of inputs are presented in (d), (h), (l)and we observe that RMSE deceases with increasing the number of inputs. Last row ((m)-(o)) is devoted to represent the effect for different group of inputs. Feeding of low degree nodes during training give better prediction than that of with the higher-degree and intermediate-degree group.  }
	\label{Fig:RMSE_diff}
\end{figure*}
\begin{figure*}
	\begin{center}
		{\includegraphics[width= 1
			\textwidth]{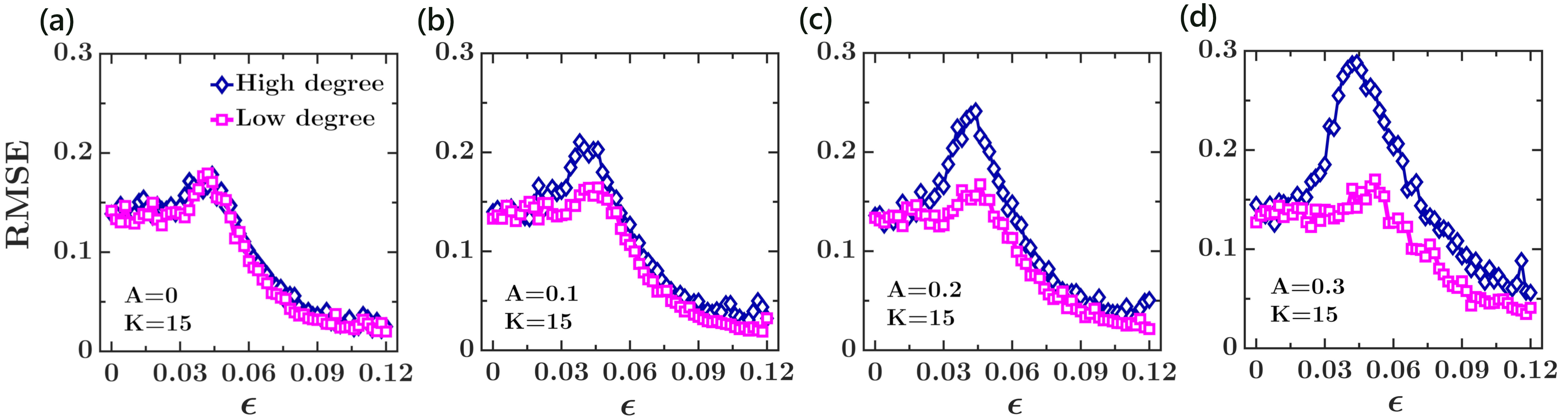}}
	\end{center}
	\caption{\textbf{RMSE for high (blue) and low (magenta) degree input nodes at different level of assortative network.} (a)A=0, (b)A=0.1, (c)A=0.2 and (d)A=0.3 are considered, and we observe the difference between RMSE for the groups of higher and lower degree nodes becomes larger with increasing assortative index. Therefore, using low degree nodes are found to be effective while training.}
	\label{Fig:RMSE}
\end{figure*}
Now, we consider assortativity mixing, i.e, a bias of a node in favor of attaching to the nodes of similar characteristics. In Fig.~\ref{Fig:RMSE_diff} we consider a degree correlated network with $A=0.3$. Here, we also choose different group of input nodes to examine the effect of degree correlation over the node selection process. The upper, middle and  lower row represent the results for high (degree$>15$), intermediate ($6\le$degree $\le15$) and low (degree$<6$) degree inputs respectively. Same as the previous figure, we use the colors black and green to represent $R$ corresponding to the original and machine generated data. Similar to the disassortative case, we find better predictions with increasing values of $K$ for each group of inputs. The last column representing RMSE, apparently verifies this result (see Fig.~\ref{Fig:RMSE_diff}(d),(h),(l)). But, here an interesting result is observed depending on the degree of input nodes. Now, we take the first column into our consideration (see Figs.~\ref{Fig:RMSE_diff} (a), (e), (i)). Here, $K=5$. We find that the training with high degree nodes do not produce the desired result. In comparison with the high degree nodes, intermediate nodes give better prediction, and the low degree nodes give more accurate result than intermediate or high degree inputs. In Fig.~\ref{Fig:RMSE_diff}(m), we plot RMSE, and observe a small variation among the results corresponding to different degree inputs. We use blue, green and yellow colors for high, intermediate and low degree inputs respectively. The discrepancy in predicted results becomes more clear for increasing values of $K$, i.e. for $K=15$ and $K=25$. RMSE for low degree input nodes remains considerably low in comparison with the high and intermediate groups (see Figs.~\ref{Fig:RMSE_diff}(n), (o)). Therefore, we notice that {\it{in an assortative network, if the low degree nodes are used to train the machine, ESN gives better results comparing to the other nodes.}} We discuss the reason behind this phenomenon on later section.

\subsection{Effect of different degree inputs at different assortativity level}

We now explore an overall scenario about the effect of assortativity index of the original system on ESN predicted results in Fig.~\ref{Fig:RMSE}. We fix $K=15$ and use the groups of high (degree $>15$) and low (degree$<6$) degree nodes for training. Here in Fig.\ref{Fig:RMSE}, initially we consider an uncorrelated network $(A=0)$ and gradually increase the degree correlation up to a certain level $(A=0.3)$ to find out the interconnection between assortativity index of the original system and RMSE regarding different groups of input nodes. Here, we plot RMSE corresponding to the considered groups of training nodes for different values of $A$. Blue and magenta colors are used for high and low degree inputs respectively. In Fig.~\ref{Fig:RMSE}(a), we plot the RMSE between the synchronization order parameters of original and machine generated data for $A=0$, and observe that the machine prediction is almost similar for two separate inputs. In Fig.~\ref{Fig:RMSE}(b)-(d), we plot the same for $A=0.1$, $A=0.2$ and $A=0.3$. Interestingly, examining this figure, we find that training with higher degree nodes fails to give the desired prediction for a degree correlated network. Here, the lower degree input nodes can give comparatively better results. A difference in the values of RMSE arises for two distinct group of inputs, and apparently this difference gradually enlarges with increasing assortativity parameter of the original system. Therefore, if we consider a degree correlated system, the low degree training nodes help the machine to make more accurate prediction.  

\begin{figure}
	\begin{center}
		{\includegraphics[width= 0.5
			\textwidth]{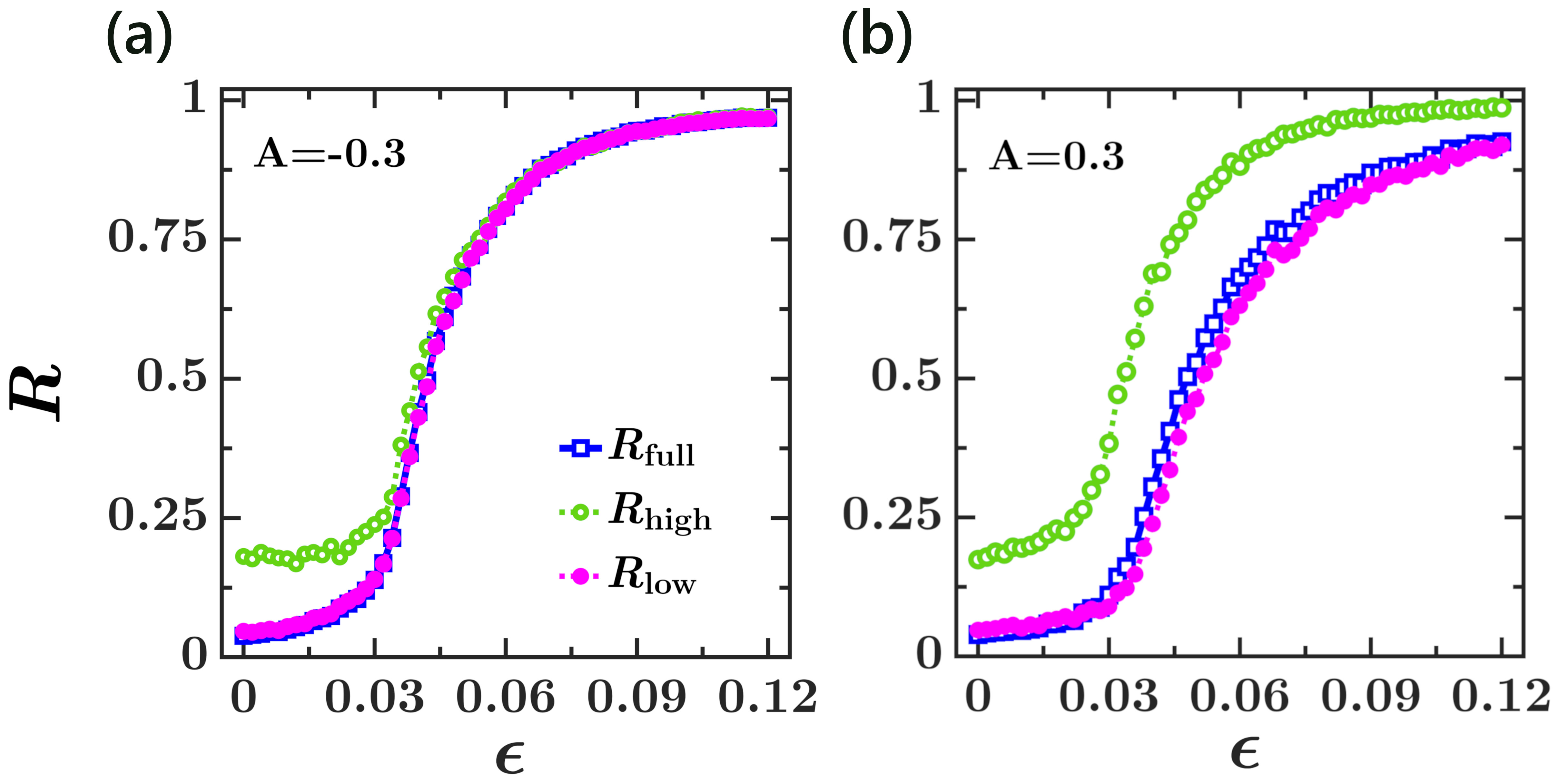}}
	\end{center}
	\caption{\textbf{ The synchronization order parameter $R$ with respect to $\epsilon$ for the whole network and two separate group of nodes of the original system} (a) Disassortative $(A=-0.3)$ and (b) assortative $(A=0.3)$ networks are considered. Blue, green and magenta colors are used for the whole system, and the group of high and low degree nodes respectively.}
	\label{Fig:group}
\end{figure}
\subsection{Mechanism}
In this section, we try to explore the reason behind the fact that why using lower degree nodes as inputs to ESN results better prediction in a highly assortative network. Furthermore, how this effect fades down with decreasing degree correlation in network topology of the original system. In Fig.\ref{Fig:RMSE}, we choose the training nodes from two distinct groups depending on their degree. We try to investigate the reason behind such discrepancy in the obtained results  depending on the input node selection. Since, based on degree of individual nodes increasing assortativity restructures the local connectivity of the network, and similar degree nodes are tending to connect with each other, here we explore the collective synchronization behavior of the original system. The path of the global synchronization order parameters of these two separate groups along with the whole system are depicted in Fig.~\ref{Fig:group}. Here, blue, green and magenta colors we use for the whole system, group of  high degree (degree$>15$) and low degree (degree$<6$) nodes respectively. \par
In Fig.~\ref{Fig:group}(a), we observe the paths to the synchronized state for a disassortative network $(A=-0.3)$. Enough connections between high and low degree nodes help to synchronize the whole system at a time. The synchronization paths are almost same for all the separate groups and the whole network. Therefore, any choice of inputs nodes does not have any non-identical effect on the results. However, in Fig.~\ref{Fig:group}(b), for highly assortative network $(A=0.3)$, we observe that different groups follow different paths to synchronize. The group of high degree nodes make a cluster and synchronizes quickly in comparison with the whole network. That might be a reason for not getting the desired prediction when we train the machine with high degree nodes. On the other hand, group of low degree nodes follows almost the same path as the whole network. Therefore, for a degree correlated network, training with low degree nodes gives a better prediction of synchronization level of the whole network.

\section{Conclusion}
\label{sec:conclusion}
In this study, we attempted to predict the burst synchronization of a neuronal system, coupled through a heterogeneous scale-free network using ESN based machine learning approach. We train the machine with the help of the time series of a few nodes and let the machine to predict the series of the remaining nodes of the system. Interestingly, ESN can not precisely predict the amplitudes of time series of each nodes, instead the onset of bursts as well as the mean field can be predicted quite well by ESN. Since, assortativity appears as a crucial characteristic for a heterogeneous network, we investigated the effect of this characteristic on the ESN predicted results. During the training session, we choose the inputs having different degree, and interestingly, different group of input nodes are found to be effective in the prediction process. In a disassortative network, no such dissimilarity is observed for different group of inputs (depending on degree). However, for a degree correlated network, the lower degree input nodes seem to be very effective. If we train ESN using the dynamics of low degree nodes, in comparison with the higher degree inputs, machine can correctly predict route of the global phase synchronization order parameter approaching synchronized state. We also analyzed the possible reason behind this result from the synchronization behavior of original system. As the lower degree nodes are very large in number, dominating the collective synchronized regime of the whole system for an assortative network, at this case low degree nodes are able to give a preferred result.\par
This investigation might be useful in providing practical insights in several fields. As ESN is a model independent technique, we might use this result in any area where only the information of some lower degree nodes is available for a heterogeneous system. At this point, if the concern is to get an idea about the collective behavior of the whole system, ESN can easily build up a desired prediction. From neuroscience viewpoint, the burst synchronized activity in some regions of the human cortex is related to human memory and cognition. So,our findings may be very useful to understand the mechanism of  human memory retrieval process.  \par

\section*{Acknowledgements}
M.R. is financially supported by University Grant Commission, Government of India.  A.S. is funded by the Center of Advanced Systems Understanding (CASUS), which is financed by Germany's Federal Ministry of Education and Research (BMBF) and by the Saxon Ministry for Science, Culture and Tourism (SMWK) with tax funds on the basis of the budget approved by the Saxon State Parliament. C.H. is supported by DST-INSPIRE-
Faculty grant (code No. IFA17-PH193).

\bibliography{References}

\begin{thebibliography}{77}%
\makeatletter
\providecommand \@ifxundefined [1]{%
 \@ifx{#1\undefined}
}%
\providecommand \@ifnum [1]{%
 \ifnum #1\expandafter \@firstoftwo
 \else \expandafter \@secondoftwo
 \fi
}%
\providecommand \@ifx [1]{%
 \ifx #1\expandafter \@firstoftwo
 \else \expandafter \@secondoftwo
 \fi
}%
\providecommand \natexlab [1]{#1}%
\providecommand \enquote  [1]{``#1''}%
\providecommand \bibnamefont  [1]{#1}%
\providecommand \bibfnamefont [1]{#1}%
\providecommand \citenamefont [1]{#1}%
\providecommand \href@noop [0]{\@secondoftwo}%
\providecommand \href [0]{\begingroup \@sanitize@url \@href}%
\providecommand \@href[1]{\@@startlink{#1}\@@href}%
\providecommand \@@href[1]{\endgroup#1\@@endlink}%
\providecommand \@sanitize@url [0]{\catcode `\\12\catcode `\$12\catcode
  `\&12\catcode `\#12\catcode `\^12\catcode `\_12\catcode `\%12\relax}%
\providecommand \@@startlink[1]{}%
\providecommand \@@endlink[0]{}%
\providecommand \url  [0]{\begingroup\@sanitize@url \@url }%
\providecommand \@url [1]{\endgroup\@href {#1}{\urlprefix }}%
\providecommand \urlprefix  [0]{URL }%
\providecommand \Eprint [0]{\href }%
\providecommand \doibase [0]{http://dx.doi.org/}%
\providecommand \selectlanguage [0]{\@gobble}%
\providecommand \bibinfo  [0]{\@secondoftwo}%
\providecommand \bibfield  [0]{\@secondoftwo}%
\providecommand \translation [1]{[#1]}%
\providecommand \BibitemOpen [0]{}%
\providecommand \bibitemStop [0]{}%
\providecommand \bibitemNoStop [0]{.\EOS\space}%
\providecommand \EOS [0]{\spacefactor3000\relax}%
\providecommand \BibitemShut  [1]{\csname bibitem#1\endcsname}%
\let\auto@bib@innerbib\@empty
\bibitem [{\citenamefont {Jaeger}(2001)}]{jaeger2001echo}%
  \BibitemOpen
  \bibfield  {author} {\bibinfo {author} {\bibfnamefont {H.}~\bibnamefont
  {Jaeger}},\ }\href@noop {} {\bibfield  {journal} {\bibinfo  {journal} {Bonn,
  Germany: German National Research Center for Information Technology GMD
  Technical Report}\ }\textbf {\bibinfo {volume} {148}},\ \bibinfo {pages} {13}
  (\bibinfo {year} {2001})}\BibitemShut {NoStop}%
\bibitem [{\citenamefont {Jaeger}\ and\ \citenamefont
  {Haas}(2004)}]{jaeger2004harnessing}%
  \BibitemOpen
  \bibfield  {author} {\bibinfo {author} {\bibfnamefont {H.}~\bibnamefont
  {Jaeger}}\ and\ \bibinfo {author} {\bibfnamefont {H.}~\bibnamefont {Haas}},\
  }\href@noop {} {\bibfield  {journal} {\bibinfo  {journal} {Science}\ }\textbf
  {\bibinfo {volume} {304}},\ \bibinfo {pages} {78} (\bibinfo {year}
  {2004})}\BibitemShut {NoStop}%
\bibitem [{\citenamefont {Luko{\v{s}}evi{\v{c}}ius}\ and\ \citenamefont
  {Jaeger}(2009)}]{lukovsevivcius2009reservoir}%
  \BibitemOpen
  \bibfield  {author} {\bibinfo {author} {\bibfnamefont {M.}~\bibnamefont
  {Luko{\v{s}}evi{\v{c}}ius}}\ and\ \bibinfo {author} {\bibfnamefont
  {H.}~\bibnamefont {Jaeger}},\ }\href@noop {} {\bibfield  {journal} {\bibinfo
  {journal} {Comput. Sci. Rev.}\ }\textbf {\bibinfo {volume} {3}},\ \bibinfo
  {pages} {127} (\bibinfo {year} {2009})}\BibitemShut {NoStop}%
\bibitem [{\citenamefont {Rodan}\ and\ \citenamefont
  {Tino}(2010)}]{rodan2010minimum}%
  \BibitemOpen
  \bibfield  {author} {\bibinfo {author} {\bibfnamefont {A.}~\bibnamefont
  {Rodan}}\ and\ \bibinfo {author} {\bibfnamefont {P.}~\bibnamefont {Tino}},\
  }\href@noop {} {\bibfield  {journal} {\bibinfo  {journal} {IEEE T. Neural
  Netw.}\ }\textbf {\bibinfo {volume} {22}},\ \bibinfo {pages} {131} (\bibinfo
  {year} {2010})}\BibitemShut {NoStop}%
\bibitem [{\citenamefont {Farka{\v{s}}}\ \emph {et~al.}(2016)\citenamefont
  {Farka{\v{s}}}, \citenamefont {Bos{\'a}k},\ and\ \citenamefont
  {Gergel'}}]{farkavs2016computational}%
  \BibitemOpen
  \bibfield  {author} {\bibinfo {author} {\bibfnamefont {I.}~\bibnamefont
  {Farka{\v{s}}}}, \bibinfo {author} {\bibfnamefont {R.}~\bibnamefont
  {Bos{\'a}k}}, \ and\ \bibinfo {author} {\bibfnamefont {P.}~\bibnamefont
  {Gergel'}},\ }\href@noop {} {\bibfield  {journal} {\bibinfo  {journal}
  {Neural Netw.}\ }\textbf {\bibinfo {volume} {83}},\ \bibinfo {pages} {109}
  (\bibinfo {year} {2016})}\BibitemShut {NoStop}%
\bibitem [{\citenamefont {Koryakin}\ \emph {et~al.}(2012)\citenamefont
  {Koryakin}, \citenamefont {Lohmann},\ and\ \citenamefont
  {Butz}}]{koryakin2012balanced}%
  \BibitemOpen
  \bibfield  {author} {\bibinfo {author} {\bibfnamefont {D.}~\bibnamefont
  {Koryakin}}, \bibinfo {author} {\bibfnamefont {J.}~\bibnamefont {Lohmann}}, \
  and\ \bibinfo {author} {\bibfnamefont {M.~V.}\ \bibnamefont {Butz}},\
  }\href@noop {} {\bibfield  {journal} {\bibinfo  {journal} {Neural Netw.}\
  }\textbf {\bibinfo {volume} {36}},\ \bibinfo {pages} {35} (\bibinfo {year}
  {2012})}\BibitemShut {NoStop}%
\bibitem [{\citenamefont {Tanaka}\ \emph {et~al.}(2019)\citenamefont {Tanaka},
  \citenamefont {Yamane}, \citenamefont {H{\'e}roux}, \citenamefont {Nakane},
  \citenamefont {Kanazawa}, \citenamefont {Takeda}, \citenamefont {Numata},
  \citenamefont {Nakano},\ and\ \citenamefont {Hirose}}]{tanaka2019recent}%
  \BibitemOpen
  \bibfield  {author} {\bibinfo {author} {\bibfnamefont {G.}~\bibnamefont
  {Tanaka}}, \bibinfo {author} {\bibfnamefont {T.}~\bibnamefont {Yamane}},
  \bibinfo {author} {\bibfnamefont {J.~B.}\ \bibnamefont {H{\'e}roux}},
  \bibinfo {author} {\bibfnamefont {R.}~\bibnamefont {Nakane}}, \bibinfo
  {author} {\bibfnamefont {N.}~\bibnamefont {Kanazawa}}, \bibinfo {author}
  {\bibfnamefont {S.}~\bibnamefont {Takeda}}, \bibinfo {author} {\bibfnamefont
  {H.}~\bibnamefont {Numata}}, \bibinfo {author} {\bibfnamefont
  {D.}~\bibnamefont {Nakano}}, \ and\ \bibinfo {author} {\bibfnamefont
  {A.}~\bibnamefont {Hirose}},\ }\href@noop {} {\bibfield  {journal} {\bibinfo
  {journal} {Neural Netw.}\ }\textbf {\bibinfo {volume} {115}},\ \bibinfo
  {pages} {100} (\bibinfo {year} {2019})}\BibitemShut {NoStop}%
\bibitem [{\citenamefont
  {Luko{\v{s}}evi{\v{c}}ius}(2012)}]{lukovsevivcius2012practical}%
  \BibitemOpen
  \bibfield  {author} {\bibinfo {author} {\bibfnamefont {M.}~\bibnamefont
  {Luko{\v{s}}evi{\v{c}}ius}},\ }in\ \href@noop {} {\emph {\bibinfo {booktitle}
  {Neural networks: Tricks of the trade}}}\ (\bibinfo  {publisher} {Springer},\
  \bibinfo {year} {2012})\ pp.\ \bibinfo {pages} {659--686}\BibitemShut
  {NoStop}%
\bibitem [{\citenamefont {Kim}\ and\ \citenamefont
  {Jeong}(2019)}]{kim2019decoding}%
  \BibitemOpen
  \bibfield  {author} {\bibinfo {author} {\bibfnamefont {H.-H.}\ \bibnamefont
  {Kim}}\ and\ \bibinfo {author} {\bibfnamefont {J.}~\bibnamefont {Jeong}},\
  }\href@noop {} {\bibfield  {journal} {\bibinfo  {journal} {Comput. Biol.
  Med.}\ }\textbf {\bibinfo {volume} {110}},\ \bibinfo {pages} {254} (\bibinfo
  {year} {2019})}\BibitemShut {NoStop}%
\bibitem [{\citenamefont {Hramov}\ \emph {et~al.}(2021)\citenamefont {Hramov},
  \citenamefont {Maksimenko},\ and\ \citenamefont
  {Pisarchik}}]{hramov2021physical}%
  \BibitemOpen
  \bibfield  {author} {\bibinfo {author} {\bibfnamefont {A.~E.}\ \bibnamefont
  {Hramov}}, \bibinfo {author} {\bibfnamefont {V.~A.}\ \bibnamefont
  {Maksimenko}}, \ and\ \bibinfo {author} {\bibfnamefont {A.~N.}\ \bibnamefont
  {Pisarchik}},\ }\href@noop {} {\bibfield  {journal} {\bibinfo  {journal}
  {Phys. Rep.}\ }\textbf {\bibinfo {volume} {918}},\ \bibinfo {pages} {1}
  (\bibinfo {year} {2021})}\BibitemShut {NoStop}%
\bibitem [{\citenamefont {Skowronski}\ and\ \citenamefont
  {Harris}(2007)}]{skowronski2007automatic}%
  \BibitemOpen
  \bibfield  {author} {\bibinfo {author} {\bibfnamefont {M.~D.}\ \bibnamefont
  {Skowronski}}\ and\ \bibinfo {author} {\bibfnamefont {J.~G.}\ \bibnamefont
  {Harris}},\ }\href@noop {} {\bibfield  {journal} {\bibinfo  {journal} {Neural
  netw.}\ }\textbf {\bibinfo {volume} {20}},\ \bibinfo {pages} {414} (\bibinfo
  {year} {2007})}\BibitemShut {NoStop}%
\bibitem [{\citenamefont {Hinaut}\ and\ \citenamefont
  {Dominey}(2013)}]{hinaut2013real}%
  \BibitemOpen
  \bibfield  {author} {\bibinfo {author} {\bibfnamefont {X.}~\bibnamefont
  {Hinaut}}\ and\ \bibinfo {author} {\bibfnamefont {P.~F.}\ \bibnamefont
  {Dominey}},\ }\href@noop {} {\bibfield  {journal} {\bibinfo  {journal} {PloS
  One}\ }\textbf {\bibinfo {volume} {8}},\ \bibinfo {pages} {e52946} (\bibinfo
  {year} {2013})}\BibitemShut {NoStop}%
\bibitem [{\citenamefont {Chessa}\ \emph {et~al.}(2014)\citenamefont {Chessa},
  \citenamefont {Gallicchio}, \citenamefont {Guzman},\ and\ \citenamefont
  {Micheli}}]{chessa2014robot}%
  \BibitemOpen
  \bibfield  {author} {\bibinfo {author} {\bibfnamefont {S.}~\bibnamefont
  {Chessa}}, \bibinfo {author} {\bibfnamefont {C.}~\bibnamefont {Gallicchio}},
  \bibinfo {author} {\bibfnamefont {R.}~\bibnamefont {Guzman}}, \ and\ \bibinfo
  {author} {\bibfnamefont {A.}~\bibnamefont {Micheli}},\ }in\ \href@noop {}
  {\emph {\bibinfo {booktitle} {Recent Advances of Neural Network Models and
  Applications}}}\ (\bibinfo  {publisher} {Springer},\ \bibinfo {year} {2014})\
  pp.\ \bibinfo {pages} {147--154}\BibitemShut {NoStop}%
\bibitem [{\citenamefont {Lin}\ \emph {et~al.}(2009)\citenamefont {Lin},
  \citenamefont {Yang},\ and\ \citenamefont {Song}}]{lin2009short}%
  \BibitemOpen
  \bibfield  {author} {\bibinfo {author} {\bibfnamefont {X.}~\bibnamefont
  {Lin}}, \bibinfo {author} {\bibfnamefont {Z.}~\bibnamefont {Yang}}, \ and\
  \bibinfo {author} {\bibfnamefont {Y.}~\bibnamefont {Song}},\ }\href@noop {}
  {\bibfield  {journal} {\bibinfo  {journal} {Expert Syst. Appl.}\ }\textbf
  {\bibinfo {volume} {36}},\ \bibinfo {pages} {7313} (\bibinfo {year}
  {2009})}\BibitemShut {NoStop}%
\bibitem [{\citenamefont {Banerjee}\ \emph {et~al.}(2019)\citenamefont
  {Banerjee}, \citenamefont {Pathak}, \citenamefont {Roy}, \citenamefont
  {Restrepo},\ and\ \citenamefont {Ott}}]{banerjee2019using}%
  \BibitemOpen
  \bibfield  {author} {\bibinfo {author} {\bibfnamefont {A.}~\bibnamefont
  {Banerjee}}, \bibinfo {author} {\bibfnamefont {J.}~\bibnamefont {Pathak}},
  \bibinfo {author} {\bibfnamefont {R.}~\bibnamefont {Roy}}, \bibinfo {author}
  {\bibfnamefont {J.~G.}\ \bibnamefont {Restrepo}}, \ and\ \bibinfo {author}
  {\bibfnamefont {E.}~\bibnamefont {Ott}},\ }\href@noop {} {\bibfield
  {journal} {\bibinfo  {journal} {Chaos}\ }\textbf {\bibinfo {volume} {29}},\
  \bibinfo {pages} {121104} (\bibinfo {year} {2019})}\BibitemShut {NoStop}%
\bibitem [{\citenamefont {Banerjee}\ \emph {et~al.}(2021)\citenamefont
  {Banerjee}, \citenamefont {Hart}, \citenamefont {Roy},\ and\ \citenamefont
  {Ott}}]{banerjee2021machine}%
  \BibitemOpen
  \bibfield  {author} {\bibinfo {author} {\bibfnamefont {A.}~\bibnamefont
  {Banerjee}}, \bibinfo {author} {\bibfnamefont {J.~D.}\ \bibnamefont {Hart}},
  \bibinfo {author} {\bibfnamefont {R.}~\bibnamefont {Roy}}, \ and\ \bibinfo
  {author} {\bibfnamefont {E.}~\bibnamefont {Ott}},\ }\href@noop {} {\bibfield
  {journal} {\bibinfo  {journal} {Phys. Rev. X}\ }\textbf {\bibinfo {volume}
  {11}},\ \bibinfo {pages} {031014} (\bibinfo {year} {2021})}\BibitemShut
  {NoStop}%
\bibitem [{\citenamefont {Panday}\ \emph {et~al.}(2021)\citenamefont {Panday},
  \citenamefont {Lee}, \citenamefont {Dutta},\ and\ \citenamefont
  {Jalan}}]{panday2021machine}%
  \BibitemOpen
  \bibfield  {author} {\bibinfo {author} {\bibfnamefont {A.}~\bibnamefont
  {Panday}}, \bibinfo {author} {\bibfnamefont {W.~S.}\ \bibnamefont {Lee}},
  \bibinfo {author} {\bibfnamefont {S.}~\bibnamefont {Dutta}}, \ and\ \bibinfo
  {author} {\bibfnamefont {S.}~\bibnamefont {Jalan}},\ }\href@noop {}
  {\bibfield  {journal} {\bibinfo  {journal} {Chaos}\ }\textbf {\bibinfo
  {volume} {31}},\ \bibinfo {pages} {031106} (\bibinfo {year}
  {2021})}\BibitemShut {NoStop}%
\bibitem [{\citenamefont {Ghosh}\ \emph {et~al.}(2021)\citenamefont {Ghosh},
  \citenamefont {Senapati}, \citenamefont {Mishra}, \citenamefont
  {Chattopadhyay}, \citenamefont {Dana}, \citenamefont {Hens},\ and\
  \citenamefont {Ghosh}}]{ghosh2021reservoir}%
  \BibitemOpen
  \bibfield  {author} {\bibinfo {author} {\bibfnamefont {S.}~\bibnamefont
  {Ghosh}}, \bibinfo {author} {\bibfnamefont {A.}~\bibnamefont {Senapati}},
  \bibinfo {author} {\bibfnamefont {A.}~\bibnamefont {Mishra}}, \bibinfo
  {author} {\bibfnamefont {J.}~\bibnamefont {Chattopadhyay}}, \bibinfo {author}
  {\bibfnamefont {S.~K.}\ \bibnamefont {Dana}}, \bibinfo {author}
  {\bibfnamefont {C.}~\bibnamefont {Hens}}, \ and\ \bibinfo {author}
  {\bibfnamefont {D.}~\bibnamefont {Ghosh}},\ }\href@noop {} {\bibfield
  {journal} {\bibinfo  {journal} {Phys. Rev. E}\ }\textbf {\bibinfo {volume}
  {104}},\ \bibinfo {pages} {014308} (\bibinfo {year} {2021})}\BibitemShut
  {NoStop}%
\bibitem [{\citenamefont {Qiao}\ \emph {et~al.}(2016)\citenamefont {Qiao},
  \citenamefont {Li}, \citenamefont {Han},\ and\ \citenamefont
  {Li}}]{qiao2016growing}%
  \BibitemOpen
  \bibfield  {author} {\bibinfo {author} {\bibfnamefont {J.}~\bibnamefont
  {Qiao}}, \bibinfo {author} {\bibfnamefont {F.}~\bibnamefont {Li}}, \bibinfo
  {author} {\bibfnamefont {H.}~\bibnamefont {Han}}, \ and\ \bibinfo {author}
  {\bibfnamefont {W.}~\bibnamefont {Li}},\ }\href@noop {} {\bibfield  {journal}
  {\bibinfo  {journal} {IEEE Trans. Neural Netw. Learn. Syst.}\ }\textbf
  {\bibinfo {volume} {28}},\ \bibinfo {pages} {391} (\bibinfo {year}
  {2016})}\BibitemShut {NoStop}%
\bibitem [{\citenamefont {Cui}\ \emph {et~al.}(2014)\citenamefont {Cui},
  \citenamefont {Feng}, \citenamefont {Chai}, \citenamefont {Liu},\ and\
  \citenamefont {Liu}}]{cui2014effect}%
  \BibitemOpen
  \bibfield  {author} {\bibinfo {author} {\bibfnamefont {H.}~\bibnamefont
  {Cui}}, \bibinfo {author} {\bibfnamefont {C.}~\bibnamefont {Feng}}, \bibinfo
  {author} {\bibfnamefont {Y.}~\bibnamefont {Chai}}, \bibinfo {author}
  {\bibfnamefont {R.~P.}\ \bibnamefont {Liu}}, \ and\ \bibinfo {author}
  {\bibfnamefont {Y.}~\bibnamefont {Liu}},\ }\href@noop {} {\bibfield
  {journal} {\bibinfo  {journal} {Neural Netw.}\ }\textbf {\bibinfo {volume}
  {57}},\ \bibinfo {pages} {141} (\bibinfo {year} {2014})}\BibitemShut
  {NoStop}%
\bibitem [{\citenamefont {Venayagamoorthy}\ and\ \citenamefont
  {Shishir}(2009)}]{venayagamoorthy2009effects}%
  \BibitemOpen
  \bibfield  {author} {\bibinfo {author} {\bibfnamefont {G.~K.}\ \bibnamefont
  {Venayagamoorthy}}\ and\ \bibinfo {author} {\bibfnamefont {B.}~\bibnamefont
  {Shishir}},\ }\href@noop {} {\bibfield  {journal} {\bibinfo  {journal}
  {Neural Netw.}\ }\textbf {\bibinfo {volume} {22}},\ \bibinfo {pages} {861}
  (\bibinfo {year} {2009})}\BibitemShut {NoStop}%
\bibitem [{\citenamefont {B{\"u}hlmann}\ and\ \citenamefont
  {Yu}(2003)}]{buhlmann2003boosting}%
  \BibitemOpen
  \bibfield  {author} {\bibinfo {author} {\bibfnamefont {P.}~\bibnamefont
  {B{\"u}hlmann}}\ and\ \bibinfo {author} {\bibfnamefont {B.}~\bibnamefont
  {Yu}},\ }\href@noop {} {\bibfield  {journal} {\bibinfo  {journal} {J Am Stat
  Assoc.}\ }\textbf {\bibinfo {volume} {98}},\ \bibinfo {pages} {324} (\bibinfo
  {year} {2003})}\BibitemShut {NoStop}%
\bibitem [{\citenamefont {Platt}\ \emph {et~al.}(2021)\citenamefont {Platt},
  \citenamefont {Wong}, \citenamefont {Clark}, \citenamefont {Penny},\ and\
  \citenamefont {Abarbanel}}]{platt2021forecasting}%
  \BibitemOpen
  \bibfield  {author} {\bibinfo {author} {\bibfnamefont {J.~A.}\ \bibnamefont
  {Platt}}, \bibinfo {author} {\bibfnamefont {A.}~\bibnamefont {Wong}},
  \bibinfo {author} {\bibfnamefont {R.}~\bibnamefont {Clark}}, \bibinfo
  {author} {\bibfnamefont {S.~G.}\ \bibnamefont {Penny}}, \ and\ \bibinfo
  {author} {\bibfnamefont {H.~D.}\ \bibnamefont {Abarbanel}},\ }\href@noop {}
  {\bibfield  {journal} {\bibinfo  {journal} {arXiv preprint arXiv:2102.08930}\
  } (\bibinfo {year} {2021})}\BibitemShut {NoStop}%
\bibitem [{\citenamefont {Griffith}\ \emph {et~al.}(2019)\citenamefont
  {Griffith}, \citenamefont {Pomerance},\ and\ \citenamefont
  {Gauthier}}]{griffith2019forecasting}%
  \BibitemOpen
  \bibfield  {author} {\bibinfo {author} {\bibfnamefont {A.}~\bibnamefont
  {Griffith}}, \bibinfo {author} {\bibfnamefont {A.}~\bibnamefont {Pomerance}},
  \ and\ \bibinfo {author} {\bibfnamefont {D.~J.}\ \bibnamefont {Gauthier}},\
  }\href@noop {} {\bibfield  {journal} {\bibinfo  {journal} {Chaos}\ }\textbf
  {\bibinfo {volume} {29}},\ \bibinfo {pages} {123108} (\bibinfo {year}
  {2019})}\BibitemShut {NoStop}%
\bibitem [{\citenamefont {Lu}\ \emph {et~al.}(2018)\citenamefont {Lu},
  \citenamefont {Hunt},\ and\ \citenamefont {Ott}}]{lu2018attractor}%
  \BibitemOpen
  \bibfield  {author} {\bibinfo {author} {\bibfnamefont {Z.}~\bibnamefont
  {Lu}}, \bibinfo {author} {\bibfnamefont {B.~R.}\ \bibnamefont {Hunt}}, \ and\
  \bibinfo {author} {\bibfnamefont {E.}~\bibnamefont {Ott}},\ }\href@noop {}
  {\bibfield  {journal} {\bibinfo  {journal} {Chaos}\ }\textbf {\bibinfo
  {volume} {28}},\ \bibinfo {pages} {061104} (\bibinfo {year}
  {2018})}\BibitemShut {NoStop}%
\bibitem [{\citenamefont {Kawai}\ \emph {et~al.}(2019)\citenamefont {Kawai},
  \citenamefont {Park},\ and\ \citenamefont {Asada}}]{kawai2019small}%
  \BibitemOpen
  \bibfield  {author} {\bibinfo {author} {\bibfnamefont {Y.}~\bibnamefont
  {Kawai}}, \bibinfo {author} {\bibfnamefont {J.}~\bibnamefont {Park}}, \ and\
  \bibinfo {author} {\bibfnamefont {M.}~\bibnamefont {Asada}},\ }\href@noop {}
  {\bibfield  {journal} {\bibinfo  {journal} {Neural Netw.}\ }\textbf {\bibinfo
  {volume} {112}},\ \bibinfo {pages} {15} (\bibinfo {year} {2019})}\BibitemShut
  {NoStop}%
\bibitem [{\citenamefont {Haluszczynski}\ and\ \citenamefont
  {Räth}(2019)}]{Alexander2019}%
  \BibitemOpen
  \bibfield  {author} {\bibinfo {author} {\bibfnamefont {A.}~\bibnamefont
  {Haluszczynski}}\ and\ \bibinfo {author} {\bibfnamefont {C.}~\bibnamefont
  {Räth}},\ }\href@noop {} {\bibfield  {journal} {\bibinfo  {journal} {Chaos}\
  }\textbf {\bibinfo {volume} {29}},\ \bibinfo {pages} {103143} (\bibinfo
  {year} {2019})}\BibitemShut {NoStop}%
\bibitem [{\citenamefont {Lymburn}\ \emph
  {et~al.}(2019{\natexlab{a}})\citenamefont {Lymburn}, \citenamefont {Khor},
  \citenamefont {Stemler}, \citenamefont {Corr{\^e}a}, \citenamefont {Small},\
  and\ \citenamefont {J{\"u}ngling}}]{lymburn2019consistency}%
  \BibitemOpen
  \bibfield  {author} {\bibinfo {author} {\bibfnamefont {T.}~\bibnamefont
  {Lymburn}}, \bibinfo {author} {\bibfnamefont {A.}~\bibnamefont {Khor}},
  \bibinfo {author} {\bibfnamefont {T.}~\bibnamefont {Stemler}}, \bibinfo
  {author} {\bibfnamefont {D.~C.}\ \bibnamefont {Corr{\^e}a}}, \bibinfo
  {author} {\bibfnamefont {M.}~\bibnamefont {Small}}, \ and\ \bibinfo {author}
  {\bibfnamefont {T.}~\bibnamefont {J{\"u}ngling}},\ }\href@noop {} {\bibfield
  {journal} {\bibinfo  {journal} {Chaos}\ }\textbf {\bibinfo {volume} {29}},\
  \bibinfo {pages} {023118} (\bibinfo {year} {2019}{\natexlab{a}})}\BibitemShut
  {NoStop}%
\bibitem [{\citenamefont {Carroll}\ and\ \citenamefont
  {Pecora}(2019)}]{carroll2019network}%
  \BibitemOpen
  \bibfield  {author} {\bibinfo {author} {\bibfnamefont {T.~L.}\ \bibnamefont
  {Carroll}}\ and\ \bibinfo {author} {\bibfnamefont {L.~M.}\ \bibnamefont
  {Pecora}},\ }\href@noop {} {\bibfield  {journal} {\bibinfo  {journal}
  {Chaos}\ }\textbf {\bibinfo {volume} {29}},\ \bibinfo {pages} {083130}
  (\bibinfo {year} {2019})}\BibitemShut {NoStop}%
\bibitem [{\citenamefont {Shirin}\ \emph {et~al.}(2019)\citenamefont {Shirin},
  \citenamefont {Klickstein},\ and\ \citenamefont
  {Sorrentino}}]{shirin2019stability}%
  \BibitemOpen
  \bibfield  {author} {\bibinfo {author} {\bibfnamefont {A.}~\bibnamefont
  {Shirin}}, \bibinfo {author} {\bibfnamefont {I.~S.}\ \bibnamefont
  {Klickstein}}, \ and\ \bibinfo {author} {\bibfnamefont {F.}~\bibnamefont
  {Sorrentino}},\ }\href@noop {} {\bibfield  {journal} {\bibinfo  {journal}
  {Chaos}\ }\textbf {\bibinfo {volume} {29}},\ \bibinfo {pages} {103147}
  (\bibinfo {year} {2019})}\BibitemShut {NoStop}%
\bibitem [{\citenamefont {Kantz}\ and\ \citenamefont
  {Schreiber}(2004)}]{kantz2004nonlinear}%
  \BibitemOpen
  \bibfield  {author} {\bibinfo {author} {\bibfnamefont {H.}~\bibnamefont
  {Kantz}}\ and\ \bibinfo {author} {\bibfnamefont {T.}~\bibnamefont
  {Schreiber}},\ }\href@noop {} {\emph {\bibinfo {title} {Nonlinear time series
  analysis}}},\ Vol.~\bibinfo {volume} {7}\ (\bibinfo  {publisher} {Cambridge
  university press},\ \bibinfo {year} {2004})\BibitemShut {NoStop}%
\bibitem [{\citenamefont {Parlitz}\ and\ \citenamefont
  {Merkwirth}(2000)}]{parlitz2000prediction}%
  \BibitemOpen
  \bibfield  {author} {\bibinfo {author} {\bibfnamefont {U.}~\bibnamefont
  {Parlitz}}\ and\ \bibinfo {author} {\bibfnamefont {C.}~\bibnamefont
  {Merkwirth}},\ }\href@noop {} {\bibfield  {journal} {\bibinfo  {journal}
  {Phys. Rev. Lett.}\ }\textbf {\bibinfo {volume} {84}},\ \bibinfo {pages}
  {1890} (\bibinfo {year} {2000})}\BibitemShut {NoStop}%
\bibitem [{\citenamefont {Pathak}\ \emph {et~al.}(2018)\citenamefont {Pathak},
  \citenamefont {Hunt}, \citenamefont {Girvan}, \citenamefont {Lu},\ and\
  \citenamefont {Ott}}]{pathak2018model}%
  \BibitemOpen
  \bibfield  {author} {\bibinfo {author} {\bibfnamefont {J.}~\bibnamefont
  {Pathak}}, \bibinfo {author} {\bibfnamefont {B.}~\bibnamefont {Hunt}},
  \bibinfo {author} {\bibfnamefont {M.}~\bibnamefont {Girvan}}, \bibinfo
  {author} {\bibfnamefont {Z.}~\bibnamefont {Lu}}, \ and\ \bibinfo {author}
  {\bibfnamefont {E.}~\bibnamefont {Ott}},\ }\href@noop {} {\bibfield
  {journal} {\bibinfo  {journal} {Phys. Rev. Lett.}\ }\textbf {\bibinfo
  {volume} {120}},\ \bibinfo {pages} {024102} (\bibinfo {year}
  {2018})}\BibitemShut {NoStop}%
\bibitem [{\citenamefont {Weng}\ \emph {et~al.}(2019)\citenamefont {Weng},
  \citenamefont {Yang}, \citenamefont {Gu}, \citenamefont {Zhang},\ and\
  \citenamefont {Small}}]{weng2019synchronization}%
  \BibitemOpen
  \bibfield  {author} {\bibinfo {author} {\bibfnamefont {T.}~\bibnamefont
  {Weng}}, \bibinfo {author} {\bibfnamefont {H.}~\bibnamefont {Yang}}, \bibinfo
  {author} {\bibfnamefont {C.}~\bibnamefont {Gu}}, \bibinfo {author}
  {\bibfnamefont {J.}~\bibnamefont {Zhang}}, \ and\ \bibinfo {author}
  {\bibfnamefont {M.}~\bibnamefont {Small}},\ }\href@noop {} {\bibfield
  {journal} {\bibinfo  {journal} {Phys. Rev. E}\ }\textbf {\bibinfo {volume}
  {99}},\ \bibinfo {pages} {042203} (\bibinfo {year} {2019})}\BibitemShut
  {NoStop}%
\bibitem [{\citenamefont {Borra}\ \emph {et~al.}(2020)\citenamefont {Borra},
  \citenamefont {Vulpiani},\ and\ \citenamefont
  {Cencini}}]{borra2020effective}%
  \BibitemOpen
  \bibfield  {author} {\bibinfo {author} {\bibfnamefont {F.}~\bibnamefont
  {Borra}}, \bibinfo {author} {\bibfnamefont {A.}~\bibnamefont {Vulpiani}}, \
  and\ \bibinfo {author} {\bibfnamefont {M.}~\bibnamefont {Cencini}},\
  }\href@noop {} {\bibfield  {journal} {\bibinfo  {journal} {Phys. Rev. E}\
  }\textbf {\bibinfo {volume} {102}},\ \bibinfo {pages} {052203} (\bibinfo
  {year} {2020})}\BibitemShut {NoStop}%
\bibitem [{\citenamefont {Maslennikov}\ and\ \citenamefont
  {Nekorkin}(2019)}]{maslennikov2019collective}%
  \BibitemOpen
  \bibfield  {author} {\bibinfo {author} {\bibfnamefont {O.~V.}\ \bibnamefont
  {Maslennikov}}\ and\ \bibinfo {author} {\bibfnamefont {V.~I.}\ \bibnamefont
  {Nekorkin}},\ }\href@noop {} {\bibfield  {journal} {\bibinfo  {journal}
  {Chaos}\ }\textbf {\bibinfo {volume} {29}},\ \bibinfo {pages} {103126}
  (\bibinfo {year} {2019})}\BibitemShut {NoStop}%
\bibitem [{\citenamefont {Zhang}\ \emph {et~al.}(2020)\citenamefont {Zhang},
  \citenamefont {Jiang}, \citenamefont {Qu},\ and\ \citenamefont
  {Lai}}]{zhang2020predicting}%
  \BibitemOpen
  \bibfield  {author} {\bibinfo {author} {\bibfnamefont {C.}~\bibnamefont
  {Zhang}}, \bibinfo {author} {\bibfnamefont {J.}~\bibnamefont {Jiang}},
  \bibinfo {author} {\bibfnamefont {S.-X.}\ \bibnamefont {Qu}}, \ and\ \bibinfo
  {author} {\bibfnamefont {Y.-C.}\ \bibnamefont {Lai}},\ }\href@noop {}
  {\bibfield  {journal} {\bibinfo  {journal} {Chaos}\ }\textbf {\bibinfo
  {volume} {30}},\ \bibinfo {pages} {083114} (\bibinfo {year}
  {2020})}\BibitemShut {NoStop}%
\bibitem [{\citenamefont {Zimmermann}\ and\ \citenamefont
  {Parlitz}(2018)}]{zimmermann2018observing}%
  \BibitemOpen
  \bibfield  {author} {\bibinfo {author} {\bibfnamefont {R.~S.}\ \bibnamefont
  {Zimmermann}}\ and\ \bibinfo {author} {\bibfnamefont {U.}~\bibnamefont
  {Parlitz}},\ }\href@noop {} {\bibfield  {journal} {\bibinfo  {journal}
  {Chaos}\ }\textbf {\bibinfo {volume} {28}},\ \bibinfo {pages} {043118}
  (\bibinfo {year} {2018})}\BibitemShut {NoStop}%
\bibitem [{\citenamefont {Pathak}\ \emph {et~al.}(2017)\citenamefont {Pathak},
  \citenamefont {Lu}, \citenamefont {Hunt}, \citenamefont {Girvan},\ and\
  \citenamefont {Ott}}]{pathak2017using}%
  \BibitemOpen
  \bibfield  {author} {\bibinfo {author} {\bibfnamefont {J.}~\bibnamefont
  {Pathak}}, \bibinfo {author} {\bibfnamefont {Z.}~\bibnamefont {Lu}}, \bibinfo
  {author} {\bibfnamefont {B.~R.}\ \bibnamefont {Hunt}}, \bibinfo {author}
  {\bibfnamefont {M.}~\bibnamefont {Girvan}}, \ and\ \bibinfo {author}
  {\bibfnamefont {E.}~\bibnamefont {Ott}},\ }\href@noop {} {\bibfield
  {journal} {\bibinfo  {journal} {Chaos}\ }\textbf {\bibinfo {volume} {27}},\
  \bibinfo {pages} {121102} (\bibinfo {year} {2017})}\BibitemShut {NoStop}%
\bibitem [{\citenamefont {Carroll}(2018)}]{carroll2018using}%
  \BibitemOpen
  \bibfield  {author} {\bibinfo {author} {\bibfnamefont {T.~L.}\ \bibnamefont
  {Carroll}},\ }\href@noop {} {\bibfield  {journal} {\bibinfo  {journal} {Phys.
  Rev. E}\ }\textbf {\bibinfo {volume} {98}},\ \bibinfo {pages} {052209}
  (\bibinfo {year} {2018})}\BibitemShut {NoStop}%
\bibitem [{\citenamefont {Kong}\ \emph {et~al.}(2021)\citenamefont {Kong},
  \citenamefont {Fan}, \citenamefont {Grebogi},\ and\ \citenamefont
  {Lai}}]{kong2021machine}%
  \BibitemOpen
  \bibfield  {author} {\bibinfo {author} {\bibfnamefont {L.-W.}\ \bibnamefont
  {Kong}}, \bibinfo {author} {\bibfnamefont {H.-W.}\ \bibnamefont {Fan}},
  \bibinfo {author} {\bibfnamefont {C.}~\bibnamefont {Grebogi}}, \ and\
  \bibinfo {author} {\bibfnamefont {Y.-C.}\ \bibnamefont {Lai}},\ }\href@noop
  {} {\bibfield  {journal} {\bibinfo  {journal} {Phys. Rev. Res.}\ }\textbf
  {\bibinfo {volume} {3}},\ \bibinfo {pages} {013090} (\bibinfo {year}
  {2021})}\BibitemShut {NoStop}%
\bibitem [{\citenamefont {Mandal}\ and\ \citenamefont
  {Shrimali}(2021)}]{mandal2021achieving}%
  \BibitemOpen
  \bibfield  {author} {\bibinfo {author} {\bibfnamefont {S.}~\bibnamefont
  {Mandal}}\ and\ \bibinfo {author} {\bibfnamefont {M.~D.}\ \bibnamefont
  {Shrimali}},\ }\href@noop {} {\bibfield  {journal} {\bibinfo  {journal}
  {Chaos}\ }\textbf {\bibinfo {volume} {31}},\ \bibinfo {pages} {031101}
  (\bibinfo {year} {2021})}\BibitemShut {NoStop}%
\bibitem [{\citenamefont {Pikovsky}\ \emph {et~al.}(2003)\citenamefont
  {Pikovsky}, \citenamefont {Kurths}, \citenamefont {Rosenblum},\ and\
  \citenamefont {Kurths}}]{pikovsky2003synchronization}%
  \BibitemOpen
  \bibfield  {author} {\bibinfo {author} {\bibfnamefont {A.}~\bibnamefont
  {Pikovsky}}, \bibinfo {author} {\bibfnamefont {J.}~\bibnamefont {Kurths}},
  \bibinfo {author} {\bibfnamefont {M.}~\bibnamefont {Rosenblum}}, \ and\
  \bibinfo {author} {\bibfnamefont {J.}~\bibnamefont {Kurths}},\ }\href@noop {}
  {\emph {\bibinfo {title} {Synchronization: a universal concept in nonlinear
  sciences}}},\ \bibinfo {number} {12}\ (\bibinfo  {publisher} {Cambridge
  university press},\ \bibinfo {year} {2003})\BibitemShut {NoStop}%
\bibitem [{\citenamefont {Arenas}\ \emph {et~al.}(2008)\citenamefont {Arenas},
  \citenamefont {D{\'\i}az-Guilera}, \citenamefont {Kurths}, \citenamefont
  {Moreno},\ and\ \citenamefont {Zhou}}]{arenas2008synchronization}%
  \BibitemOpen
  \bibfield  {author} {\bibinfo {author} {\bibfnamefont {A.}~\bibnamefont
  {Arenas}}, \bibinfo {author} {\bibfnamefont {A.}~\bibnamefont
  {D{\'\i}az-Guilera}}, \bibinfo {author} {\bibfnamefont {J.}~\bibnamefont
  {Kurths}}, \bibinfo {author} {\bibfnamefont {Y.}~\bibnamefont {Moreno}}, \
  and\ \bibinfo {author} {\bibfnamefont {C.}~\bibnamefont {Zhou}},\ }\href@noop
  {} {\bibfield  {journal} {\bibinfo  {journal} {Phys. Rep.}\ }\textbf
  {\bibinfo {volume} {469}},\ \bibinfo {pages} {93} (\bibinfo {year}
  {2008})}\BibitemShut {NoStop}%
\bibitem [{\citenamefont {Koseska}\ \emph {et~al.}(2013)\citenamefont
  {Koseska}, \citenamefont {Volkov},\ and\ \citenamefont
  {Kurths}}]{koseska2013oscillation}%
  \BibitemOpen
  \bibfield  {author} {\bibinfo {author} {\bibfnamefont {A.}~\bibnamefont
  {Koseska}}, \bibinfo {author} {\bibfnamefont {E.}~\bibnamefont {Volkov}}, \
  and\ \bibinfo {author} {\bibfnamefont {J.}~\bibnamefont {Kurths}},\
  }\href@noop {} {\bibfield  {journal} {\bibinfo  {journal} {Phys. Rep.}\
  }\textbf {\bibinfo {volume} {531}},\ \bibinfo {pages} {173} (\bibinfo {year}
  {2013})}\BibitemShut {NoStop}%
\bibitem [{\citenamefont {Saxena}\ \emph {et~al.}(2012)\citenamefont {Saxena},
  \citenamefont {Prasad},\ and\ \citenamefont
  {Ramaswamy}}]{saxena2012amplitude}%
  \BibitemOpen
  \bibfield  {author} {\bibinfo {author} {\bibfnamefont {G.}~\bibnamefont
  {Saxena}}, \bibinfo {author} {\bibfnamefont {A.}~\bibnamefont {Prasad}}, \
  and\ \bibinfo {author} {\bibfnamefont {R.}~\bibnamefont {Ramaswamy}},\
  }\href@noop {} {\bibfield  {journal} {\bibinfo  {journal} {Phys. Rep.}\
  }\textbf {\bibinfo {volume} {521}},\ \bibinfo {pages} {205} (\bibinfo {year}
  {2012})}\BibitemShut {NoStop}%
\bibitem [{\citenamefont {Parastesh}\ \emph {et~al.}(2021)\citenamefont
  {Parastesh}, \citenamefont {Jafari}, \citenamefont {Azarnoush}, \citenamefont
  {Shahriari}, \citenamefont {Wang}, \citenamefont {Boccaletti},\ and\
  \citenamefont {Perc}}]{parastesh2020chimeras}%
  \BibitemOpen
  \bibfield  {author} {\bibinfo {author} {\bibfnamefont {F.}~\bibnamefont
  {Parastesh}}, \bibinfo {author} {\bibfnamefont {S.}~\bibnamefont {Jafari}},
  \bibinfo {author} {\bibfnamefont {H.}~\bibnamefont {Azarnoush}}, \bibinfo
  {author} {\bibfnamefont {Z.}~\bibnamefont {Shahriari}}, \bibinfo {author}
  {\bibfnamefont {Z.}~\bibnamefont {Wang}}, \bibinfo {author} {\bibfnamefont
  {S.}~\bibnamefont {Boccaletti}}, \ and\ \bibinfo {author} {\bibfnamefont
  {M.}~\bibnamefont {Perc}},\ }\href@noop {} {\bibfield  {journal} {\bibinfo
  {journal} {Phys. Rep.}\ }\textbf {\bibinfo {volume} {898}},\ \bibinfo {pages}
  {1} (\bibinfo {year} {2021})}\BibitemShut {NoStop}%
\bibitem [{\citenamefont {Xiao}\ \emph {et~al.}(2021)\citenamefont {Xiao},
  \citenamefont {Kong}, \citenamefont {Sun},\ and\ \citenamefont
  {Lai}}]{xiao2021predicting}%
  \BibitemOpen
  \bibfield  {author} {\bibinfo {author} {\bibfnamefont {R.}~\bibnamefont
  {Xiao}}, \bibinfo {author} {\bibfnamefont {L.-W.}\ \bibnamefont {Kong}},
  \bibinfo {author} {\bibfnamefont {Z.-K.}\ \bibnamefont {Sun}}, \ and\
  \bibinfo {author} {\bibfnamefont {Y.-C.}\ \bibnamefont {Lai}},\ }\href@noop
  {} {\bibfield  {journal} {\bibinfo  {journal} {Phys. Rev. E}\ }\textbf
  {\bibinfo {volume} {104}},\ \bibinfo {pages} {014205} (\bibinfo {year}
  {2021})}\BibitemShut {NoStop}%
\bibitem [{\citenamefont {Lymburn}\ \emph
  {et~al.}(2019{\natexlab{b}})\citenamefont {Lymburn}, \citenamefont {Walker},
  \citenamefont {Small},\ and\ \citenamefont
  {J{\"u}ngling}}]{lymburn2019reservoir}%
  \BibitemOpen
  \bibfield  {author} {\bibinfo {author} {\bibfnamefont {T.}~\bibnamefont
  {Lymburn}}, \bibinfo {author} {\bibfnamefont {D.~M.}\ \bibnamefont {Walker}},
  \bibinfo {author} {\bibfnamefont {M.}~\bibnamefont {Small}}, \ and\ \bibinfo
  {author} {\bibfnamefont {T.}~\bibnamefont {J{\"u}ngling}},\ }\href@noop {}
  {\bibfield  {journal} {\bibinfo  {journal} {Chaos}\ }\textbf {\bibinfo
  {volume} {29}},\ \bibinfo {pages} {093133} (\bibinfo {year}
  {2019}{\natexlab{b}})}\BibitemShut {NoStop}%
\bibitem [{\citenamefont {Ib{\'a}{\~n}ez-Soria}\ \emph
  {et~al.}(2018)\citenamefont {Ib{\'a}{\~n}ez-Soria}, \citenamefont
  {Garc{\'\i}a-Ojalvo}, \citenamefont {Soria-Frisch},\ and\ \citenamefont
  {Ruffini}}]{ibanez2018detection}%
  \BibitemOpen
  \bibfield  {author} {\bibinfo {author} {\bibfnamefont {D.}~\bibnamefont
  {Ib{\'a}{\~n}ez-Soria}}, \bibinfo {author} {\bibfnamefont {J.}~\bibnamefont
  {Garc{\'\i}a-Ojalvo}}, \bibinfo {author} {\bibfnamefont {A.}~\bibnamefont
  {Soria-Frisch}}, \ and\ \bibinfo {author} {\bibfnamefont {G.}~\bibnamefont
  {Ruffini}},\ }\href@noop {} {\bibfield  {journal} {\bibinfo  {journal}
  {Chaos}\ }\textbf {\bibinfo {volume} {28}},\ \bibinfo {pages} {033118}
  (\bibinfo {year} {2018})}\BibitemShut {NoStop}%
\bibitem [{\citenamefont {Fan}\ \emph {et~al.}(2021)\citenamefont {Fan},
  \citenamefont {Kong}, \citenamefont {Lai},\ and\ \citenamefont
  {Wang}}]{fan2021anticipating}%
  \BibitemOpen
  \bibfield  {author} {\bibinfo {author} {\bibfnamefont {H.}~\bibnamefont
  {Fan}}, \bibinfo {author} {\bibfnamefont {L.-W.}\ \bibnamefont {Kong}},
  \bibinfo {author} {\bibfnamefont {Y.-C.}\ \bibnamefont {Lai}}, \ and\
  \bibinfo {author} {\bibfnamefont {X.}~\bibnamefont {Wang}},\ }\href@noop {}
  {\bibfield  {journal} {\bibinfo  {journal} {Phys. Rev. Res.}\ }\textbf
  {\bibinfo {volume} {3}},\ \bibinfo {pages} {023237} (\bibinfo {year}
  {2021})}\BibitemShut {NoStop}%
\bibitem [{\citenamefont {Ganaie}\ \emph {et~al.}(2020)\citenamefont {Ganaie},
  \citenamefont {Ghosh}, \citenamefont {Mendola}, \citenamefont {Tanveer},\
  and\ \citenamefont {Jalan}}]{ganaie2020identification}%
  \BibitemOpen
  \bibfield  {author} {\bibinfo {author} {\bibfnamefont {M.}~\bibnamefont
  {Ganaie}}, \bibinfo {author} {\bibfnamefont {S.}~\bibnamefont {Ghosh}},
  \bibinfo {author} {\bibfnamefont {N.}~\bibnamefont {Mendola}}, \bibinfo
  {author} {\bibfnamefont {M.}~\bibnamefont {Tanveer}}, \ and\ \bibinfo
  {author} {\bibfnamefont {S.}~\bibnamefont {Jalan}},\ }\href@noop {}
  {\bibfield  {journal} {\bibinfo  {journal} {Chaos}\ }\textbf {\bibinfo
  {volume} {30}},\ \bibinfo {pages} {063128} (\bibinfo {year}
  {2020})}\BibitemShut {NoStop}%
\bibitem [{\citenamefont {Kushwaha}\ \emph {et~al.}(2021)\citenamefont
  {Kushwaha}, \citenamefont {Mendola}, \citenamefont {Ghosh}, \citenamefont
  {Kachhvah},\ and\ \citenamefont {Jalan}}]{kushwaha2021machine}%
  \BibitemOpen
  \bibfield  {author} {\bibinfo {author} {\bibfnamefont {N.}~\bibnamefont
  {Kushwaha}}, \bibinfo {author} {\bibfnamefont {N.~K.}\ \bibnamefont
  {Mendola}}, \bibinfo {author} {\bibfnamefont {S.}~\bibnamefont {Ghosh}},
  \bibinfo {author} {\bibfnamefont {A.~D.}\ \bibnamefont {Kachhvah}}, \ and\
  \bibinfo {author} {\bibfnamefont {S.}~\bibnamefont {Jalan}},\ }\href@noop {}
  {\bibfield  {journal} {\bibinfo  {journal} {Front. Phys.}\ }\textbf {\bibinfo
  {volume} {9}},\ \bibinfo {pages} {147} (\bibinfo {year} {2021})}\BibitemShut
  {NoStop}%
\bibitem [{\citenamefont {Frolov}\ \emph {et~al.}(2019)\citenamefont {Frolov},
  \citenamefont {Maksimenko}, \citenamefont {L{\"u}ttjohann}, \citenamefont
  {Koronovskii},\ and\ \citenamefont {Hramov}}]{frolov2019feed}%
  \BibitemOpen
  \bibfield  {author} {\bibinfo {author} {\bibfnamefont {N.}~\bibnamefont
  {Frolov}}, \bibinfo {author} {\bibfnamefont {V.}~\bibnamefont {Maksimenko}},
  \bibinfo {author} {\bibfnamefont {A.}~\bibnamefont {L{\"u}ttjohann}},
  \bibinfo {author} {\bibfnamefont {A.}~\bibnamefont {Koronovskii}}, \ and\
  \bibinfo {author} {\bibfnamefont {A.}~\bibnamefont {Hramov}},\ }\href@noop {}
  {\bibfield  {journal} {\bibinfo  {journal} {Chaos}\ }\textbf {\bibinfo
  {volume} {29}},\ \bibinfo {pages} {091101} (\bibinfo {year}
  {2019})}\BibitemShut {NoStop}%
\bibitem [{\citenamefont {Chowdhury}\ \emph {et~al.}(2021)\citenamefont
  {Chowdhury}, \citenamefont {Ray}, \citenamefont {Mishra},\ and\ \citenamefont
  {Ghosh}}]{chowdhury2021extreme}%
  \BibitemOpen
  \bibfield  {author} {\bibinfo {author} {\bibfnamefont {S.~N.}\ \bibnamefont
  {Chowdhury}}, \bibinfo {author} {\bibfnamefont {A.}~\bibnamefont {Ray}},
  \bibinfo {author} {\bibfnamefont {A.}~\bibnamefont {Mishra}}, \ and\ \bibinfo
  {author} {\bibfnamefont {D.}~\bibnamefont {Ghosh}},\ }\href@noop {}
  {\bibfield  {journal} {\bibinfo  {journal} {Journal of Physics: Complexity}\
  }\textbf {\bibinfo {volume} {2}},\ \bibinfo {pages} {035021} (\bibinfo {year}
  {2021})}\BibitemShut {NoStop}%
\bibitem [{\citenamefont {Ray}\ \emph {et~al.}(2021)\citenamefont {Ray},
  \citenamefont {Chakraborty},\ and\ \citenamefont {Ghosh}}]{ray2021optimized}%
  \BibitemOpen
  \bibfield  {author} {\bibinfo {author} {\bibfnamefont {A.}~\bibnamefont
  {Ray}}, \bibinfo {author} {\bibfnamefont {T.}~\bibnamefont {Chakraborty}}, \
  and\ \bibinfo {author} {\bibfnamefont {D.}~\bibnamefont {Ghosh}},\
  }\href@noop {} {\bibfield  {journal} {\bibinfo  {journal} {arXiv preprint
  arXiv:2106.08968}\ } (\bibinfo {year} {2021})}\BibitemShut {NoStop}%
\bibitem [{\citenamefont {Chakraborty}\ and\ \citenamefont
  {Ghosh}(2020)}]{chakraborty2020real}%
  \BibitemOpen
  \bibfield  {author} {\bibinfo {author} {\bibfnamefont {T.}~\bibnamefont
  {Chakraborty}}\ and\ \bibinfo {author} {\bibfnamefont {I.}~\bibnamefont
  {Ghosh}},\ }\href@noop {} {\bibfield  {journal} {\bibinfo  {journal} {Chaos,
  Solitons \& Fractals}\ }\textbf {\bibinfo {volume} {135}},\ \bibinfo {pages}
  {109850} (\bibinfo {year} {2020})}\BibitemShut {NoStop}%
\bibitem [{\citenamefont {Izhikevich}(2000)}]{izhikevich2000neural}%
  \BibitemOpen
  \bibfield  {author} {\bibinfo {author} {\bibfnamefont {E.~M.}\ \bibnamefont
  {Izhikevich}},\ }\href@noop {} {\bibfield  {journal} {\bibinfo  {journal}
  {Int J Bifurcat Chaos}\ }\textbf {\bibinfo {volume} {10}},\ \bibinfo {pages}
  {1171} (\bibinfo {year} {2000})}\BibitemShut {NoStop}%
\bibitem [{\citenamefont {Hens}\ \emph {et~al.}(2015)\citenamefont {Hens},
  \citenamefont {Pal},\ and\ \citenamefont {Dana}}]{hens2015bursting}%
  \BibitemOpen
  \bibfield  {author} {\bibinfo {author} {\bibfnamefont {C.}~\bibnamefont
  {Hens}}, \bibinfo {author} {\bibfnamefont {P.}~\bibnamefont {Pal}}, \ and\
  \bibinfo {author} {\bibfnamefont {S.~K.}\ \bibnamefont {Dana}},\ }\href@noop
  {} {\bibfield  {journal} {\bibinfo  {journal} {Phys. Rev. E}\ }\textbf
  {\bibinfo {volume} {92}},\ \bibinfo {pages} {022915} (\bibinfo {year}
  {2015})}\BibitemShut {NoStop}%
\bibitem [{\citenamefont {Mishra}\ \emph {et~al.}(2021)\citenamefont {Mishra},
  \citenamefont {Ghosh}, \citenamefont {Kumar~Dana}, \citenamefont
  {Kapitaniak},\ and\ \citenamefont {Hens}}]{mishra2021neuron}%
  \BibitemOpen
  \bibfield  {author} {\bibinfo {author} {\bibfnamefont {A.}~\bibnamefont
  {Mishra}}, \bibinfo {author} {\bibfnamefont {S.}~\bibnamefont {Ghosh}},
  \bibinfo {author} {\bibfnamefont {S.}~\bibnamefont {Kumar~Dana}}, \bibinfo
  {author} {\bibfnamefont {T.}~\bibnamefont {Kapitaniak}}, \ and\ \bibinfo
  {author} {\bibfnamefont {C.}~\bibnamefont {Hens}},\ }\href@noop {} {\bibfield
   {journal} {\bibinfo  {journal} {Chaos}\ }\textbf {\bibinfo {volume} {31}},\
  \bibinfo {pages} {052101} (\bibinfo {year} {2021})}\BibitemShut {NoStop}%
\bibitem [{\citenamefont {Roy}\ \emph {et~al.}(2021)\citenamefont {Roy},
  \citenamefont {Poria},\ and\ \citenamefont {Hens}}]{roy2021assortativity}%
  \BibitemOpen
  \bibfield  {author} {\bibinfo {author} {\bibfnamefont {M.}~\bibnamefont
  {Roy}}, \bibinfo {author} {\bibfnamefont {S.}~\bibnamefont {Poria}}, \ and\
  \bibinfo {author} {\bibfnamefont {C.}~\bibnamefont {Hens}},\ }\href@noop {}
  {\bibfield  {journal} {\bibinfo  {journal} {Phys. Rev. E}\ }\textbf {\bibinfo
  {volume} {103}},\ \bibinfo {pages} {062307} (\bibinfo {year}
  {2021})}\BibitemShut {NoStop}%
\bibitem [{\citenamefont {Rulkov}(2002)}]{rulkov2002modeling}%
  \BibitemOpen
  \bibfield  {author} {\bibinfo {author} {\bibfnamefont {N.~F.}\ \bibnamefont
  {Rulkov}},\ }\href@noop {} {\bibfield  {journal} {\bibinfo  {journal} {Phys.
  Rev. E}\ }\textbf {\bibinfo {volume} {65}},\ \bibinfo {pages} {041922}
  (\bibinfo {year} {2002})}\BibitemShut {NoStop}%
\bibitem [{\citenamefont {Ghosh}\ \emph {et~al.}(2020)\citenamefont {Ghosh},
  \citenamefont {Mondal}, \citenamefont {Ji}, \citenamefont {Mishra},
  \citenamefont {Dana}, \citenamefont {Antonopoulos},\ and\ \citenamefont
  {Hens}}]{ghosh2020emergence}%
  \BibitemOpen
  \bibfield  {author} {\bibinfo {author} {\bibfnamefont {S.}~\bibnamefont
  {Ghosh}}, \bibinfo {author} {\bibfnamefont {A.}~\bibnamefont {Mondal}},
  \bibinfo {author} {\bibfnamefont {P.}~\bibnamefont {Ji}}, \bibinfo {author}
  {\bibfnamefont {A.}~\bibnamefont {Mishra}}, \bibinfo {author} {\bibfnamefont
  {S.~K.}\ \bibnamefont {Dana}}, \bibinfo {author} {\bibfnamefont {C.~G.}\
  \bibnamefont {Antonopoulos}}, \ and\ \bibinfo {author} {\bibfnamefont
  {C.}~\bibnamefont {Hens}},\ }\href@noop {} {\bibfield  {journal} {\bibinfo
  {journal} {Front. Comput. Neurosci.}\ }\textbf {\bibinfo {volume} {14}},\
  \bibinfo {pages} {49} (\bibinfo {year} {2020})}\BibitemShut {NoStop}%
\bibitem [{\citenamefont {Saha}\ \emph {et~al.}(2020)\citenamefont {Saha},
  \citenamefont {Mishra}, \citenamefont {Ghosh}, \citenamefont {Dana},\ and\
  \citenamefont {Hens}}]{saha2020predicting}%
  \BibitemOpen
  \bibfield  {author} {\bibinfo {author} {\bibfnamefont {S.}~\bibnamefont
  {Saha}}, \bibinfo {author} {\bibfnamefont {A.}~\bibnamefont {Mishra}},
  \bibinfo {author} {\bibfnamefont {S.}~\bibnamefont {Ghosh}}, \bibinfo
  {author} {\bibfnamefont {S.~K.}\ \bibnamefont {Dana}}, \ and\ \bibinfo
  {author} {\bibfnamefont {C.}~\bibnamefont {Hens}},\ }\href@noop {} {\bibfield
   {journal} {\bibinfo  {journal} {Phys. Rev. Res.}\ }\textbf {\bibinfo
  {volume} {2}},\ \bibinfo {pages} {033338} (\bibinfo {year}
  {2020})}\BibitemShut {NoStop}%
\bibitem [{\citenamefont {Eguiluz}\ \emph {et~al.}(2005)\citenamefont
  {Eguiluz}, \citenamefont {Chialvo}, \citenamefont {Cecchi}, \citenamefont
  {Baliki},\ and\ \citenamefont {Apkarian}}]{eguiluz2005scale}%
  \BibitemOpen
  \bibfield  {author} {\bibinfo {author} {\bibfnamefont {V.~M.}\ \bibnamefont
  {Eguiluz}}, \bibinfo {author} {\bibfnamefont {D.~R.}\ \bibnamefont
  {Chialvo}}, \bibinfo {author} {\bibfnamefont {G.~A.}\ \bibnamefont {Cecchi}},
  \bibinfo {author} {\bibfnamefont {M.}~\bibnamefont {Baliki}}, \ and\ \bibinfo
  {author} {\bibfnamefont {A.~V.}\ \bibnamefont {Apkarian}},\ }\href@noop {}
  {\bibfield  {journal} {\bibinfo  {journal} {Phys. Rev. Lett.}\ }\textbf
  {\bibinfo {volume} {94}},\ \bibinfo {pages} {018102} (\bibinfo {year}
  {2005})}\BibitemShut {NoStop}%
\bibitem [{\citenamefont {Zhou}\ \emph {et~al.}(2006)\citenamefont {Zhou},
  \citenamefont {Zemanov{\'a}}, \citenamefont {Zamora}, \citenamefont
  {Hilgetag},\ and\ \citenamefont {Kurths}}]{zhou2006hierarchical}%
  \BibitemOpen
  \bibfield  {author} {\bibinfo {author} {\bibfnamefont {C.}~\bibnamefont
  {Zhou}}, \bibinfo {author} {\bibfnamefont {L.}~\bibnamefont {Zemanov{\'a}}},
  \bibinfo {author} {\bibfnamefont {G.}~\bibnamefont {Zamora}}, \bibinfo
  {author} {\bibfnamefont {C.~C.}\ \bibnamefont {Hilgetag}}, \ and\ \bibinfo
  {author} {\bibfnamefont {J.}~\bibnamefont {Kurths}},\ }\href@noop {}
  {\bibfield  {journal} {\bibinfo  {journal} {Phys. Rev. Lett.}\ }\textbf
  {\bibinfo {volume} {97}},\ \bibinfo {pages} {238103} (\bibinfo {year}
  {2006})}\BibitemShut {NoStop}%
\bibitem [{\citenamefont {Bassett}\ and\ \citenamefont
  {Bullmore}(2006)}]{bassett2006small}%
  \BibitemOpen
  \bibfield  {author} {\bibinfo {author} {\bibfnamefont {D.~S.}\ \bibnamefont
  {Bassett}}\ and\ \bibinfo {author} {\bibfnamefont {E.}~\bibnamefont
  {Bullmore}},\ }\href@noop {} {\bibfield  {journal} {\bibinfo  {journal}
  {Neuroscientist}\ }\textbf {\bibinfo {volume} {12}},\ \bibinfo {pages} {512}
  (\bibinfo {year} {2006})}\BibitemShut {NoStop}%
\bibitem [{\citenamefont {Bassett}\ \emph {et~al.}(2006)\citenamefont
  {Bassett}, \citenamefont {Meyer-Lindenberg}, \citenamefont {Achard},
  \citenamefont {Duke},\ and\ \citenamefont {Bullmore}}]{bassett2006adaptive}%
  \BibitemOpen
  \bibfield  {author} {\bibinfo {author} {\bibfnamefont {D.~S.}\ \bibnamefont
  {Bassett}}, \bibinfo {author} {\bibfnamefont {A.}~\bibnamefont
  {Meyer-Lindenberg}}, \bibinfo {author} {\bibfnamefont {S.}~\bibnamefont
  {Achard}}, \bibinfo {author} {\bibfnamefont {T.}~\bibnamefont {Duke}}, \ and\
  \bibinfo {author} {\bibfnamefont {E.}~\bibnamefont {Bullmore}},\ }\href@noop
  {} {\bibfield  {journal} {\bibinfo  {journal} {Proc. Natl. Acad. Sci.
  U.S.A.}\ }\textbf {\bibinfo {volume} {103}},\ \bibinfo {pages} {19518}
  (\bibinfo {year} {2006})}\BibitemShut {NoStop}%
\bibitem [{\citenamefont {Newman}(2002)}]{newman2002assortative}%
  \BibitemOpen
  \bibfield  {author} {\bibinfo {author} {\bibfnamefont {M.~E.}\ \bibnamefont
  {Newman}},\ }\href@noop {} {\bibfield  {journal} {\bibinfo  {journal} {Phys.
  Rev. Lett.}\ }\textbf {\bibinfo {volume} {89}},\ \bibinfo {pages} {208701}
  (\bibinfo {year} {2002})}\BibitemShut {NoStop}%
\bibitem [{\citenamefont {Newman}(2003)}]{newman2003mixing}%
  \BibitemOpen
  \bibfield  {author} {\bibinfo {author} {\bibfnamefont {M.~E.}\ \bibnamefont
  {Newman}},\ }\href@noop {} {\bibfield  {journal} {\bibinfo  {journal} {Phys.
  Rev. E}\ }\textbf {\bibinfo {volume} {67}},\ \bibinfo {pages} {026126}
  (\bibinfo {year} {2003})}\BibitemShut {NoStop}%
\bibitem [{\citenamefont {Yang}\ \emph {et~al.}(2015)\citenamefont {Yang},
  \citenamefont {Tang},\ and\ \citenamefont {Lai}}]{yang2015traffic}%
  \BibitemOpen
  \bibfield  {author} {\bibinfo {author} {\bibfnamefont {H.-X.}\ \bibnamefont
  {Yang}}, \bibinfo {author} {\bibfnamefont {M.}~\bibnamefont {Tang}}, \ and\
  \bibinfo {author} {\bibfnamefont {Y.-C.}\ \bibnamefont {Lai}},\ }\href@noop
  {} {\bibfield  {journal} {\bibinfo  {journal} {Phys. Rev. E}\ }\textbf
  {\bibinfo {volume} {91}},\ \bibinfo {pages} {062817} (\bibinfo {year}
  {2015})}\BibitemShut {NoStop}%
\bibitem [{\citenamefont {Schl{\"a}pfer}\ and\ \citenamefont
  {Buzna}(2012)}]{schlapfer2012decelerated}%
  \BibitemOpen
  \bibfield  {author} {\bibinfo {author} {\bibfnamefont {M.}~\bibnamefont
  {Schl{\"a}pfer}}\ and\ \bibinfo {author} {\bibfnamefont {L.}~\bibnamefont
  {Buzna}},\ }\href@noop {} {\bibfield  {journal} {\bibinfo  {journal} {Phys.
  Rev. E}\ }\textbf {\bibinfo {volume} {85}},\ \bibinfo {pages} {015101}
  (\bibinfo {year} {2012})}\BibitemShut {NoStop}%
\bibitem [{\citenamefont {Pastor-Satorras}\ \emph {et~al.}(2001)\citenamefont
  {Pastor-Satorras}, \citenamefont {V{\'a}zquez},\ and\ \citenamefont
  {Vespignani}}]{pastor2001dynamical}%
  \BibitemOpen
  \bibfield  {author} {\bibinfo {author} {\bibfnamefont {R.}~\bibnamefont
  {Pastor-Satorras}}, \bibinfo {author} {\bibfnamefont {A.}~\bibnamefont
  {V{\'a}zquez}}, \ and\ \bibinfo {author} {\bibfnamefont {A.}~\bibnamefont
  {Vespignani}},\ }\href@noop {} {\bibfield  {journal} {\bibinfo  {journal}
  {Phys. Rev. Lett.}\ }\textbf {\bibinfo {volume} {87}},\ \bibinfo {pages}
  {258701} (\bibinfo {year} {2001})}\BibitemShut {NoStop}%
\bibitem [{\citenamefont {Batista}\ \emph {et~al.}(2007)\citenamefont
  {Batista}, \citenamefont {Batista}, \citenamefont {De~Pontes}, \citenamefont
  {Viana},\ and\ \citenamefont {Lopes}}]{batista2007chaotic}%
  \BibitemOpen
  \bibfield  {author} {\bibinfo {author} {\bibfnamefont {C.}~\bibnamefont
  {Batista}}, \bibinfo {author} {\bibfnamefont {A.}~\bibnamefont {Batista}},
  \bibinfo {author} {\bibfnamefont {J.}~\bibnamefont {De~Pontes}}, \bibinfo
  {author} {\bibfnamefont {R.}~\bibnamefont {Viana}}, \ and\ \bibinfo {author}
  {\bibfnamefont {S.}~\bibnamefont {Lopes}},\ }\href@noop {} {\bibfield
  {journal} {\bibinfo  {journal} {Phys. Rev. E}\ }\textbf {\bibinfo {volume}
  {76}},\ \bibinfo {pages} {016218} (\bibinfo {year} {2007})}\BibitemShut
  {NoStop}%
\bibitem [{\citenamefont {Barab{\'a}si}(2009)}]{barabasi2009scale}%
  \BibitemOpen
  \bibfield  {author} {\bibinfo {author} {\bibfnamefont {A.-L.}\ \bibnamefont
  {Barab{\'a}si}},\ }\href@noop {} {\bibfield  {journal} {\bibinfo  {journal}
  {Science}\ }\textbf {\bibinfo {volume} {325}},\ \bibinfo {pages} {412}
  (\bibinfo {year} {2009})}\BibitemShut {NoStop}%
\bibitem [{\citenamefont {Kuramoto}(1991)}]{kuramoto1991collective}%
  \BibitemOpen
  \bibfield  {author} {\bibinfo {author} {\bibfnamefont {Y.}~\bibnamefont
  {Kuramoto}},\ }\href@noop {} {\bibfield  {journal} {\bibinfo  {journal}
  {Physica D}\ }\textbf {\bibinfo {volume} {50}},\ \bibinfo {pages} {15}
  (\bibinfo {year} {1991})}\BibitemShut {NoStop}%
\bibitem [{\citenamefont {Xulvi-Brunet}\ and\ \citenamefont
  {Sokolov}(2004)}]{xulvi2004reshuffling}%
  \BibitemOpen
  \bibfield  {author} {\bibinfo {author} {\bibfnamefont {R.}~\bibnamefont
  {Xulvi-Brunet}}\ and\ \bibinfo {author} {\bibfnamefont {I.~M.}\ \bibnamefont
  {Sokolov}},\ }\href@noop {} {\bibfield  {journal} {\bibinfo  {journal} {Phys.
  Rev. E}\ }\textbf {\bibinfo {volume} {70}},\ \bibinfo {pages} {066102}
  (\bibinfo {year} {2004})}\BibitemShut {NoStop}%
\end{thebibliography}%
\bibliographystyle{apsrev4-1}

\end{document}